\documentclass[aps,prb, twocolumn,floatfix,notitlepage, superscriptaddress,longbibliography]{revtex4-2}
\usepackage[utf8]{inputenc}
\usepackage{amsfonts,amsmath,amssymb,graphicx,epstopdf,verbatim}
\usepackage{rotating}
\usepackage{epstopdf}
\usepackage{xcolor}
\usepackage{natbib}
\usepackage[colorlinks]{hyperref}
\definecolor{darkred}{rgb}{0.5,0,0}
\definecolor{darkgreen}{rgb}{0,0.5,0}
\definecolor{darkblue}{rgb}{0,0,0.5}
\hypersetup{colorlinks,
linkcolor=darkred,
filecolor=darkgreen,
urlcolor=darkblue,
citecolor=darkgreen}
\usepackage{braket}
\usepackage{overpic}

\newcommand{\nep}{\operatorname{e}}

\begin{document}

\title{Spatiotemporally ordered patterns in a chain of coupled dissipative kicked rotors}

\author{Angelo Russomanno}
\affiliation{Scuola Superiore Meridionale, Università di Napoli Federico II
Largo San Marcellino 10, I-80138 Napoli, Italy and Dipartimento di Fisica, Università di Napoli Federico II, Monte Sant’Angelo, I-80126 Napoli, Italy}
\begin{abstract}
In this work we consider the dynamics of a chain of many coupled kicked rotors with dissipation. We map a rich phase diagram with many dynamical regimes. 
We focus mainly on a regime where the system shows period doubling, and forms patterns that are persistent and depend on the stroboscopic time with period double than that of the driving: The system shows a form of spatiotemporal ordering analogous to quantum Floquet time crystals. {Spatiotemporally ordered patterns can be understood by means of a linear-stability analysis that predicts an instability region that contains the spatiotemporally ordered regime. The boundary of the instability region coincides with the lower boundary of the spatiotemporally ordered regime, and the most unstable mode has length scale double than the lattice spacing, a feature that we observe in the spatiotemporally ordered patterns: Period doubling occurs both in time and space.} We 
propose an implementation of this model in an array of SQUID Josephson junctions with a pulsed time-periodic flux.
\end{abstract}
\maketitle
\section{Introduction}
In many-body dynamical systems out of equilibrium, ordered coherent patterns in space and time naturally appear, from the convective Rayleigh-B\'enard cells, to the heart beats, to the Belusov-Zabotinskii reaction, to synchronization~\cite{ball:book,strogatz:book,cross:book}. In this context there are some universal properties. One of these is the period-doubling bifurcation cascade~\cite{baba:book,gold:book}. In simple systems undergoing a periodic driving, like one-dimensional maps, one can see that increasing a parameter, the system undergoes a sequence of period doublings. At each of these transitions, a response of the system appears with period double than the regime before. As a result, one has a response with period $2^q$ times the driving, and for the parameter tending towards a finite value, $q\to\infty$ and the system becomes chaotic. This phenomenon occurs with universal scaling properties near the transition to chaos, independently of the precise choice of the map~\cite{Feigenbaum,Briggs}.

The period-doubling cascade has been observed in Nature in many contexts, from the convection rolls in water and mercury~\cite{PhysRevLett.47.243,Libchaber_1982}, to nonlinear electronic circuits~\cite{PhysRevLett.47.1349,PhysRevLett.48.714}, neurons~\cite{Bing}, and infinite-range dissipative quantum spin systems described by a mean-field theory in the thermodynamic limit~\cite{Hartmann_2017}. All these systems have in common the fact that they are nonlinear, are described by few effective variables, and undergo a periodic driving.

When the periodic driving is applied to many-body systems, the situation changes: As argued in~\cite{PhysRevA.41.1932}, the generic response is the period doubling. Responses at larger multiples of the period are possible, but only in case of special symmetries. Any spatially ordered pattern with a period $m$ times the driving with $m>2$ is doomed to be destroyed by the growth of bubbles. The only possible stable patterns are the ones with period doubling, where at each period the driving exchanges the inner and the outer of the bubble, that therefore alternatively grows and shrinks.

This result is of great importance for the recent researches on discrete time crystals in classical noisy periodically-driven systems~\cite{zhuang2021absolutely,Pizzi_2021,PhysRevE.100.060105}, where a persistent response at a multiple of the driving period appears only in the thermodynamic limit~\cite{notarella1}, and in all the examples known at this time this response occurs as a period doubling, in agreement with the results of~\cite{PhysRevA.41.1932} described above. (See also Refs.~\cite{tcrev1,tcrev2,tcrev3} for general reviews on time crystals.) 

Motivated by this framework, we aim to understand how the period-doubling cascade changes when a many-body context is considered. In order to do that, we go beyond the more usual framework of coupled maps~\cite{n1,n2,n3} and consider an array of coupled kicked rotors with dissipation, a model easy to numerically simulate, also at large sizes. Without dissipation, this model reduces to a slight generalization~\cite{PhysRevE.97.022202} of the coupled kicked rotors, showing Hamiltonian chaos~\cite{PhysRevA.40.6130,Konishi_1990,PhysRevA.44.2263,chirikov,chirikov1,PhysRevB.100.100302}. Without the coupling this model reduces to the single dissipative kicked rotor, known also as Zaslavsky map~\cite{zasla,dittrich}. This model shows a peculiar strange attractor and a very interesting dynamics~\cite{zasla}. At the onset of chaos the single-rotor model shows a behavior very similar to two parallel period doubling cascades, and we can study its fate in the case of many coupled rotors. 


We probe the system stroboscopically, that's to say at discrete times, integer multiple of the driving period, and we consider appropriate averages over random initial conditions.
Using some ``period-doubling order parameters'' inspired by the literature on time crystals~\cite{Else_2016,PhysRevB.99.104303} we see that the period-doubling cascade is washed away and the model can show essentially only period doubling in the regime of regular dynamics. In small parameter ranges there is a response at a period 4 times the driving, but it is many order of magnitude smaller than the one at period 2. So, we find that the findings of~\cite{PhysRevA.41.1932} are essentially confirmed, with the small period 4 response due probably to the fact that this model has continuous onsite variables. 

\begin{figure}
  \begin{tabular}{c}
      \includegraphics[width=80mm]{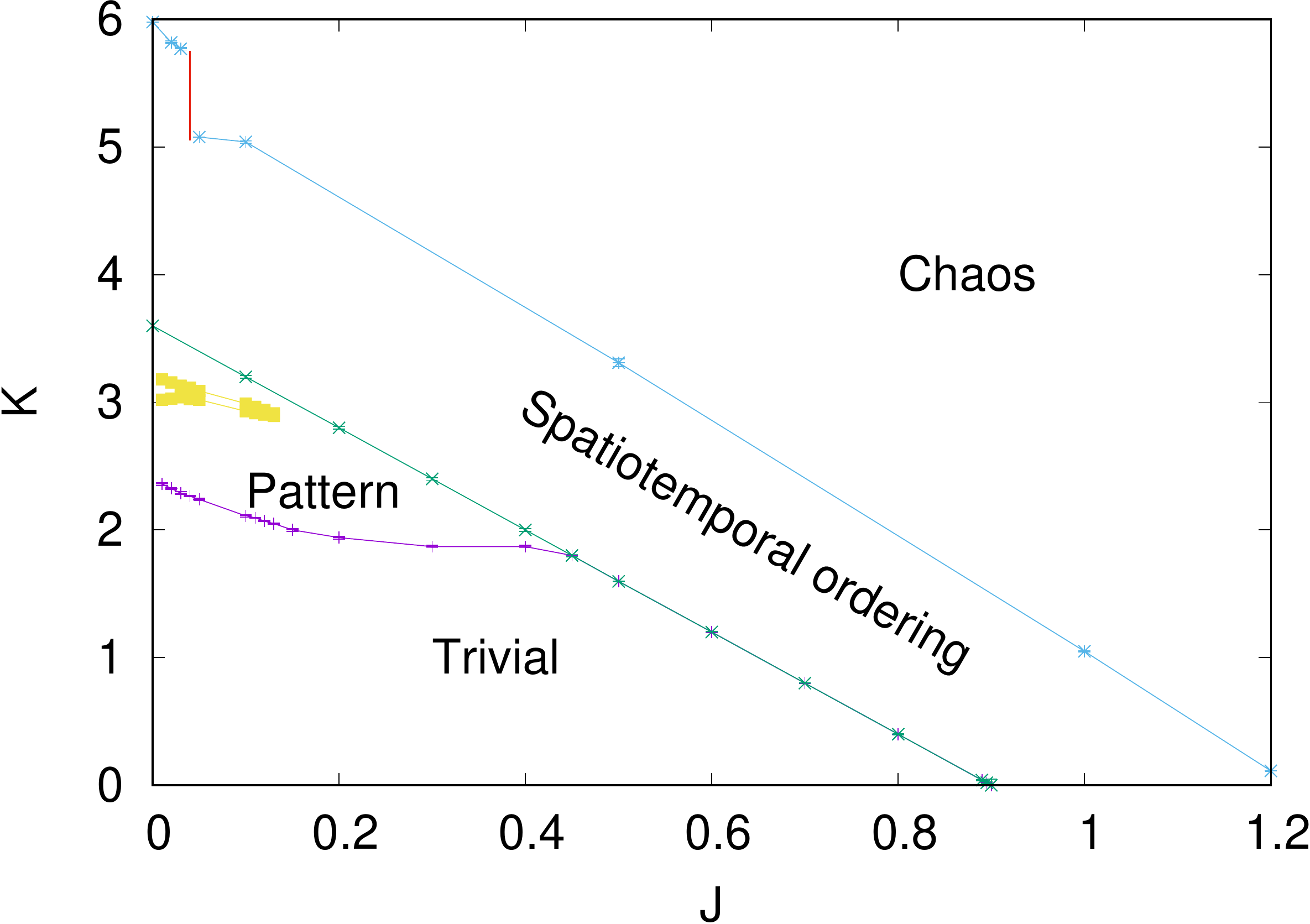}
    \end{tabular}
     \caption{Sketch of the dynamical phase diagram for $\gamma=0.8$. (For other values of $\gamma$ the situation is qualitatively similar.) We recognize the trivial regime (where relaxation to a uniform asymptotic condition occurs), the ``Pattern'' regime (where the system breaks space translation symmetry by generating persistent time-independent patterns), the ``Spatiotemporal ordered'' regime (where the persistent patterns depend on the stroboscopic time with a period double the one of the driving, breaking thereby space and time-translation symmetry), and the chaotic regime, where the dynamics is aperiodic in space and time. Inside the yellow region there is the weak patterning regime (see text for a description).}\label{phd:fig}
\end{figure}

 
Beyond the period $m$-tupling behavior, this model shows a very rich dynamical behavior, and we summarize it in the phase diagram shown in Fig.~\ref{phd:fig}. 
Anticipating a little bit, on the axes there are two parameters, $J$ and $K$, of the model (it is discussed in detail in Sec.~\ref{mode:sec}), while another parameter, $\gamma$, is kept fixed at $\gamma=0.8$.

Let us first of all focus on the regime where period doubling occurs, that in the phase diagram we term ``Spatiotemporal ordering''. This name is due to the fact that, whenever period doubling occurs, the system spontaneously organizes in space, breaking the translation symmetry and giving rise to patterns. These patterns are stable and persistent in time, change with a period twice the one of the driving, and so give rise to the period doubling. They are an example of an effect of nonlinear dynamics in spatially extended systems very common in Nature~\cite{cross:book}, and the precise form of the patterns depend on the initial state chosen.

{It is important to remark that the spatiotemporally ordered regime can be analytically understood by means of a linear stability analysis done according to the method explained in~\cite{cross:book}. We find an instability region that contains the spatiotemporally ordered regime and shares with it a boundary. Moreover, the eigenvalues in the unstable region are such that the unstable mode increases changing sign at every period, consistent with a period-doubling regime. (That sets on when nonlinear effects become strong enough to contrast this increase.) Finally, the linear-stability analysis predicts that the most unstable mode appears at momentum $k=\pi$, that's to say at a length scale double than the lattice spacing. Using the Fourier transform, we find that also the fully formed patterns have the same typical length scale, and so doubling of the time periodicity comes together with doubling of the space periodicity. This analysis is very important: From one side provides analytical predictions (of the lower boundary of the spatiotemporally ordered regime, and of the typical length scale of the patterns); From the other shows that spatiotemporally ordered patterns are true patterns in the sense of~\cite{cross:book}, that's to say instabilities with spatial organization around a uniform state, that increase and are eventually stabilized by nonlinear effects.}


Our spatiotemporal ordering has some analogy with time crystals, where time-translation symmetry breaking comes together with the breaking of an internal symmetry~\cite{Else_2016,von_Keyserlingk_2016,Khemani_2016}, but is a different phenomenon. Indeed, we see a persistent period doubling response already at finite sizes, while a true time crystal should break the space and time translation symmetry only in the thermodynamic limit. We see here an effect of nonlinear dynamics, a mechanism physically different to the quantum phase transition-like behavior involved in quantum Floquet time crystals (see for instance the discussion in~\cite{PhysRevB.99.104303}).

{There are other regimes in the phase diagram of Fig.~\ref{phd:fig}. Let us start with the ``Pattern'' one, where the system shows persistent patterns independent of the stroboscopic time. In this regime the system breaks the space translation symmetry, but not the discrete time translation symmetry. {Unfortunately we are not able to interpret these time-independent patterns in terms of a linear-stability analysis.} We see that the time-independent patterning exists only for $K$ smaller than a threshold ($K<0.45$)
and find a small region (surrounded by yellow lines in Fig.~\ref{phd:fig}), where ``weak patterning'' occurs: In a very jagged and seemingly fractal way, there are points without patterning and points with patterns with very small amplitude. The transition from time-independent patterns to spatiotemporal ordering can be seen in the typical length scale of the patterns, that suddenly drops at the boundary. This length scale in the spatiotemporally ordered case approximately coincides with the one analytically predicted using the linear-stability analysis.}


In the region labeled as ``Trivial'', the system relaxes to a uniform and time-independent condition, where all the momenta are vanishing. In the chaotic regime, in contrast, nearby trajectories in the phase space diverge exponentially from each other, and the dynamics is given by aperiodic oscillations in space and time. Here the largest Lyapunov exponent (LLE -- the measure of the rate of exponential divergence) is positive. The transition from negative (regular dynamics) to positive (chaotic dynamics) is always sharp, but along the segment marked in red, where the LLE is near to 0 and intermittently becomes slightly positive. 

Many of the transitions between the different regimes described above can be seen in the behavior of the kinetic energy per site, a quantity often considered in studies about kicked rotors~\cite{PhysRevA.40.6130,Konishi_1990,chirikov2,PhysRevE.97.022202}. In contrast with cases with Hamiltonian chaos, where this quantity increases steadily and without a bound~\cite{PhysRevA.40.6130,Konishi_1990,chirikov2,PhysRevB.100.100302}, here the kinetic energy per site reaches an asymptotic value, that does not scale with the system size. 

This is an important information for experimental realizations. Indeed, we propose an experimental realization of this model with an array of SQUID Josephson junctions with a time-periodic pulsed magnetic flux. The kinetic energy per site translates in the charging energy per site of the superconducting system, and the fact that it is bounded provides the possibility to tune the parameters so that this energy per site stays below the superconducting gap. In this way the array of Josephson junctions can keep superconductivity for long times, and be correctly described by our model.

The paper is organized as follows. In Sec.~\ref{mode:sec} we introduce the model we study. {In Sec.~\ref{stability:sec} we perform the linear stability analysis around the trivial regime, and show that there is an instability appearing exactly where the spatiotemporally ordered regime sets in. In Sec.~\ref{pt:sec} we discuss the patterns and their properties -- amplitude and typical length scale -- and relate them to the predictions of linear-stability analysis.} In Sec.~\ref{chaos:sec} we study chaos by means of the largest Lyapunov exponent, and map the boundary line of the ``Chaos'' region in Fig.~\ref{phd:fig}.  In Sec.~\ref{mtup:sec} we study the period $m$-tupling in the single- and many- rotor cases, and show that in the many-rotor model only period doubling survives. In Sec.~\ref{nono:sec} we discuss the behavior of the kinetic energy per site. In Sec.~\ref{exp:sec} we discuss how to realize our model in an array of SQUID Josephson junctions with pulsed time-periodic magnetic flux. In Sec.~\ref{conc:sec} we draw our conclusions.
\section{Model}\label{mode:sec}
We add dissipation to the many-body generalization of the kicked rotor considered in~\cite{PhysRevE.97.022202}, a slight generalization to the paradigmatic model of many-body Hamiltonian chaos theory studied in~\cite{PhysRevA.40.6130,Konishi_1990,PhysRevA.44.2263,chirikov,chirikov1,PhysRevB.100.100302}, that can be even realized experimentally with an array of bosonic Josephson junctions~\cite{trombettoni,PhysRevB.100.100302}. The purely Hamiltonian model is given by the kicked Hamiltonian
\begin{equation}\label{hamiltone:eqn}
  H(t) = \sum_{j=1}^L\left[\frac{p_j^2}{2}-\delta_1(t)\left(J\cos(\theta_j-\theta_{j+1})+K\cos(\theta_j)\right)\right]\,,
\end{equation}
where $\theta_j$, $p_j$ are pairs of canonically conjugated variables. We assume periodic boundary conditions ($\theta_{L+1}=\theta_1$), and we have defined the periodic delta function $\delta_\tau(t) \equiv \sum_{n\in\mathbb{N}}\delta(t-n\tau)$ as in~\cite{chirikov}. {The fact that $\tau=1$ is not a limitation, because $\tau$ can always be rescaled away in the definition of the couplings~\cite{chirikov2})} We see that the system is a 1-dimensional chain of rotors with kicked short-range interactions and we focus on the stroboscopic dynamics, probing the system at discrete times $t_n=n^-$, where the superscript ``-'' means that we look at the system just before the $n$-the kick has been applied. 

Writing the canonical equations of the dynamics $$\dot{\theta}_j=\partial_{p_j} H,\quad\dot{p}_j=-\partial_{\theta_j} H,$$ and using a standard analysis (essentially the integration of the second equation across the $\delta$ function -- see for instance~\cite{chirikov2,PhysRevA.40.6130,Konishi_1990,hans:book}) we see that the dynamics from $t_{n}$ to $t_{n+1}$ is described by a discrete map
\begin{align}\label{hami:eqn}
  p_j^{(n+1)} &= p_j^{(n)} - J\left[\sin(\theta_j^{(n)}-\theta_{j+1}^{(n)})+\sin(\theta_j^{(n)}-\theta_{j-1}^{(n)})\right]\nonumber\\
    &-K\sin(\theta_j^{(n)})\nonumber\\
  \theta_j^{(n+1)}&=\theta_j^{(n)} + p_j^{(n+1)}\,,
\end{align}
for $j=1,\,\ldots,\,L$, where we have written for simplicity $p_j(t_n) = p_j(n)$, $\theta_j(n) = \theta_j(t_n)$. 

\begin{figure}
  \begin{tabular}{c}
      (a) \\
      \includegraphics[width=80mm]{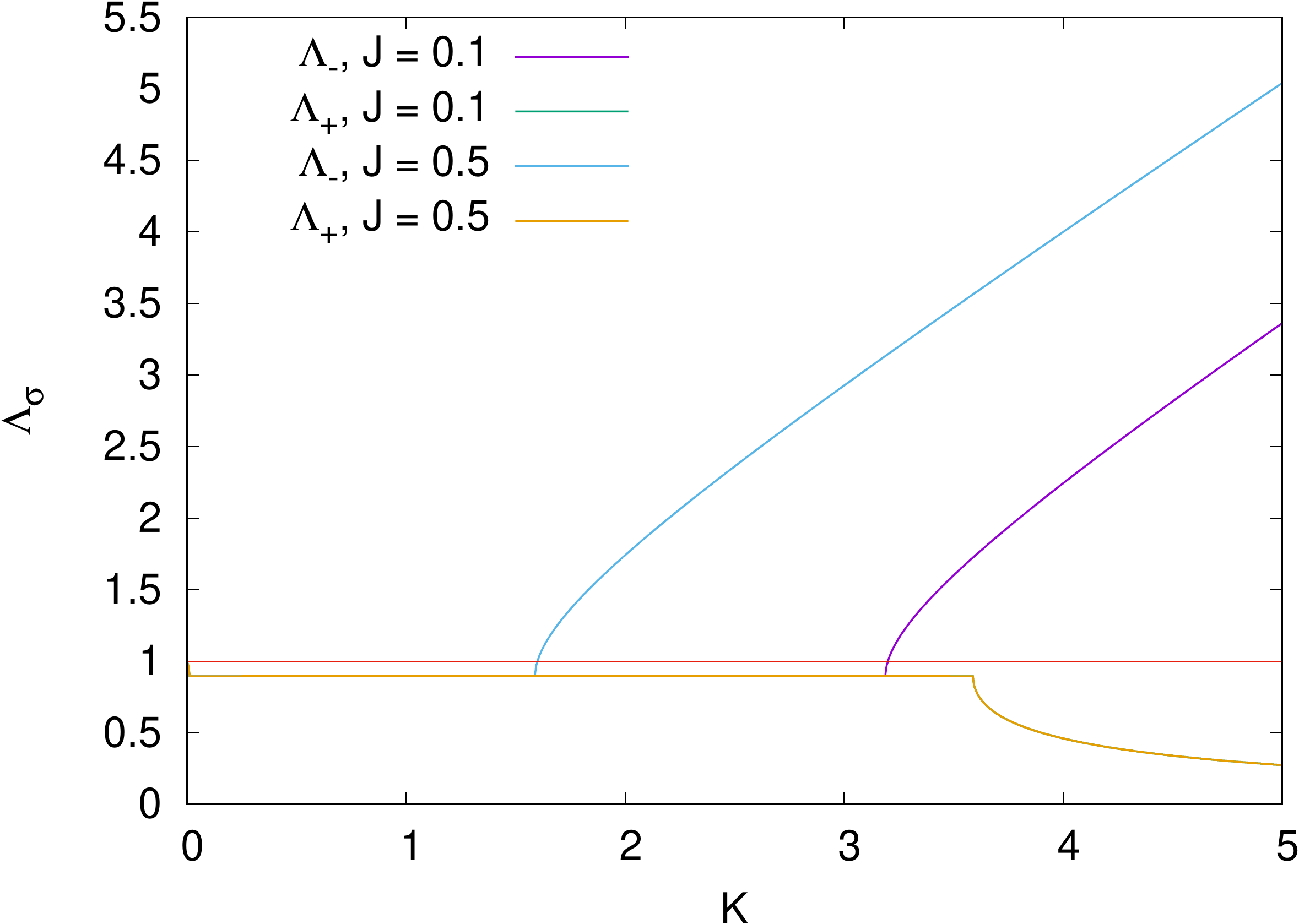}\\
      (b) \\
      \includegraphics[width=80mm]{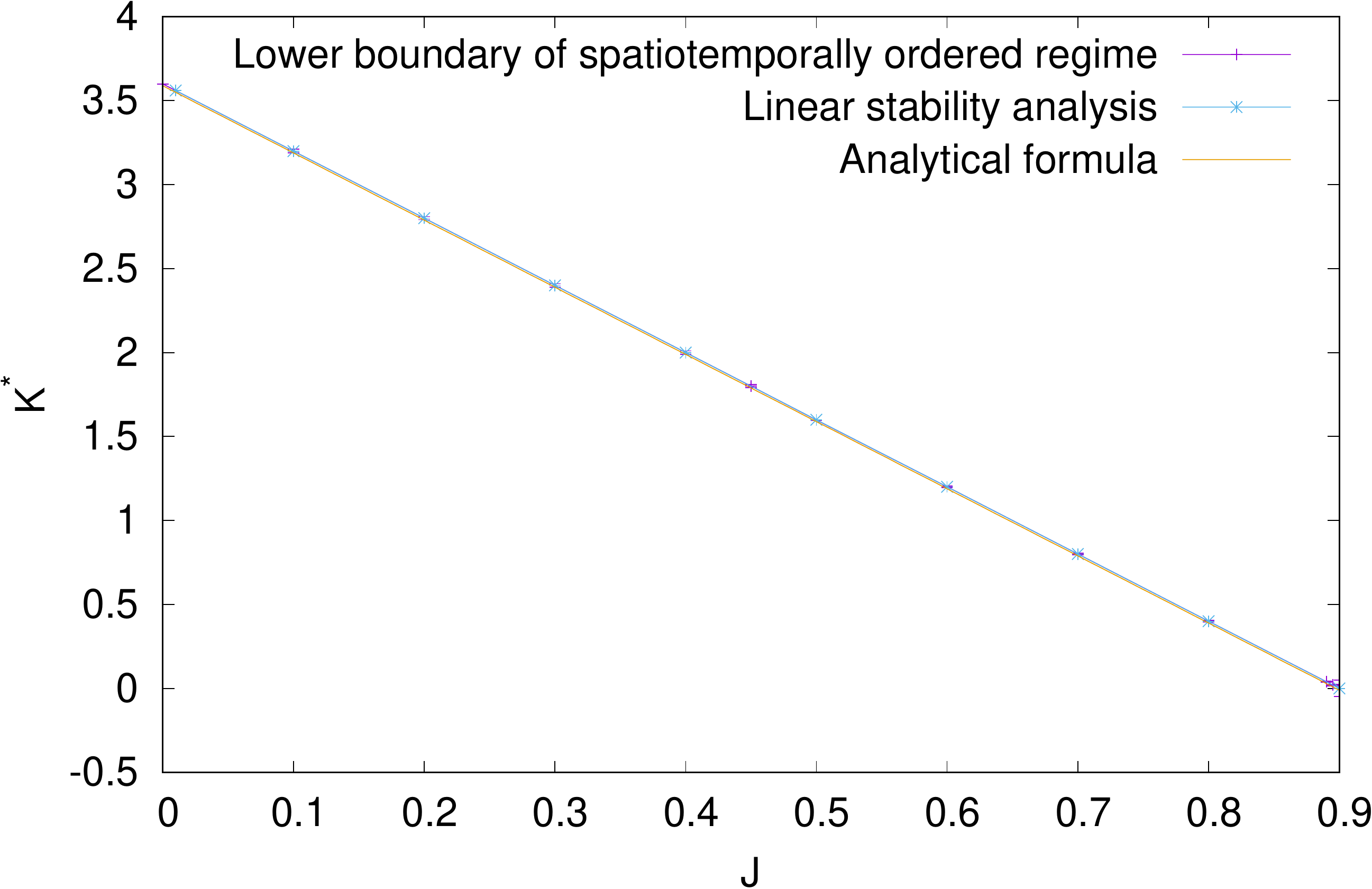}
    \end{tabular}
     \caption{(a) Maximum absolute value of the eigenvalues $\Lambda_{\sigma}$ versus $K$, for $\sigma\in\{-,\,+\}$ and two values of $J$. There is a $K^*$ such that, for $K>K^*$, $\Lambda_{-}$ becomes definitively larger than 1 (red horizontal line), marking the setting on of the instability. $\Lambda_{+}$ versus $K$ is independent of $J$ and equals 1 when $K = 0$. (b) Comparison between the instability boundary line provided by the linear stability analysis and the lower boundary line of the spatiotemporally ordered regime: They coincide perfectly and are very good approximated by the analytical formula in Eq.~\eqref{est:eqn}.}\label{patmax:fig}
\end{figure}

We add to this model a dissipation, so that between one kick and the next the momenta are damped by a factor $0<\gamma<1$. The resulting map is
\begin{align}
  p_j^{(n+1)} &= \gamma p_j^{(n)} - J\left[\sin(\theta_j^{(n)}-\theta_{j+1}^{(n)})+\sin(\theta_j^{(n)}-\theta_{j-1}^{(n)})\right]\nonumber\\
   &-K\sin(\theta_j^{(n)})\nonumber\\
  \theta_j^{(n+1)}&=\theta_j^{(n)} + p_j^{(n+1)}\,,
\end{align}
for $j=1,\,\ldots,\,L$. This is the many-body generalization of the well-known single kicked rotor with dissipation~\cite{dittrich} -- known also as Zaslavsky map~\cite{zasla} --, to which our model reduces for $L=1$
\begin{align}\label{sing:eqn}
  p^{(n+1)}&=\gamma p^{(n)} - K \sin(\theta^{(n)})\nonumber\\
  \theta^{(n+1)}&=\theta^{(n)} + p^{(n+1)}\,.
\end{align}
Here we keep fixed $\gamma=0.8$ (other choices of $\gamma$ would give a qualitatively similar phase diagram). Where not otherwise specified, we will always consider appropriate averages over random initial conditions, taking in each of them all the $p_j^{(0)}$ and the $\theta_j^{(0)}$ from a random distribution uniform in the interval $[-1,1]$.
Of this model we study chaotic properties, period $m$-tupling properties, and patterning. For the first one we use the largest Lyapunov exponent that we consider in Sec.~\ref{chaos:sec}, for the second we define an appropriate set of order parameters in Sec.~\ref{mtup:sec}, and discuss patterning in Sec.~\ref{pt:sec}. We can gain analytical insight in patterning using a linear-stability analysis, as we are going to show in the next section.
{\section{Linear stability analysis}}\label{stability:sec}
In order to understand when the patterning appears, we use the method described in~\cite{cross:book}, and linearize the dynamical equations Eq.~\eqref{sing:eqn} around the uniformly vanishing solution of Eq.~\eqref{sing:eqn}, the one $p_j^{(n)}\equiv 0$, $\theta_j^{(n)}\equiv 0$. This is the stable solution for $K$ and $J$ small enough (``Trivial'' regime), and corresponds to the trivial phase in Fig.~\ref{phd:fig}. The linearized equations are
\begin{align}
  p_j^{(n+1)} &= \gamma p_j^{(n)} - J\left[2\theta_j^{(n)}-\theta_{j-1}^{(n)}-\theta_{j+1}^{(n)}\right]-K\theta_j^{(n)}\nonumber\\
  \theta_j^{(n+1)}&=\theta_j^{(n)} + p_j^{(n+1)}\,.
\end{align}
At this point let us apply the Fourier transform $p_j^{(n)} = \sum_k p_{k}^{(n)} \nep^{ik j}$, $\theta_j^{(n)} = \sum_k \theta_k^{(n)} \nep^{ik j}$, where $k$ can take the values consistent with periodic boundary conditions, that's to say $k=\frac{2\pi n}{L}$, with $n=0,\,1,\,\ldots,\,L-1$ integer. After the Fourier transform we get
%
%
\begin{equation}
  \left(\begin{array}{c}p_k^{(n+1)}\\\theta_k^{(n+1)}\end{array}\right) = \left(\begin{array}{cc}\gamma &  - \left[2J\left(1-\cos(k)\right)+K\right] \\
                                            \gamma & 1 - \left[2J\left(1-\cos(k)\right)+K\right]\end{array}\right)\left(\begin{array}{c}p_k^{(n)}\\\theta_k^{(n)}\end{array}\right)\,.
\end{equation}
To study stability, we need to evaluate the eigenvalues of this $2\times 2$ matrix. There is instability (and pattern generation) whenever the absolute value of at least one of the eigenvalues is larger than 1. The eigenvalues $\mu$ are the solutions of the equation
$$
  \mu^2- \left\{\gamma + 1- 2J\left(1-\cos(k)\right)-K \right\}\mu + \gamma = 0\,.
$$
These solutions are explicitly
\begin{align}\label{eqig:eqn}
  &\mu_{\pm}^{(k)} = \frac{1}{2}\Big[\gamma + 1 - 2J\left(1-\cos(k)\right)-K\nonumber\\
  &\pm\sqrt{\left[\gamma + 1 - 2J\left(1-\cos(k)\right)-K\right]^2-4\gamma}\Big]\,.
\end{align}
In order to see when the uniformly vanishing solution becomes unstable, we have to look at the maximum over $k$  of the absolute values of $\mu_{\sigma}^{(k)}$. Let us call them $\Lambda_{\sigma}=\max_k|\mu_{\sigma}^{(k)}|$, with $\sigma\in\{-,\,+\}$: When at least one of these maxima becomes larger than 1 the instability sets on and the system can generate patterns. We plot some example of $\Lambda_{\pm}$ versus $K$ in Fig.~\ref{patmax:fig}(a). 

We see that  $\Lambda_{-}$ increases with $K$, and there is a critical value $K^*$ beyond which  $\Lambda_{-}$ becomes definitively larger than 1. $K^*$ therefore sets the onset of the instability. We plot $K^*$ versus $J$ in Fig.~\ref{patmax:fig}(b), and see that perfectly coincides with the lower boundary line of the period-doubling spatiotemporally ordered regime (see the phase diagram in Fig.~\ref{phd:fig} and the analysis in Sec.~\ref{double:sec}, to see how this boundary line is numerically obtained). So, the linear-stability analysis is able to quantitatively reproduce the setting on of the spatiotemporal ordering, but does not describe the regime of time-independent patterning.

When there is instability, the maximum of $|\mu_{-}^{(k)}|$ occurs always for $k_{\rm max}^{+}=\pi$, as we have numerically checked. So, $k=\pi$ is the momentum of the most unstable mode. This implies that the instability provides a pattern with a typical length scale double than the lattice step. So, we see that doubling of the time periodicity comes together with doubling of the space periodicity of the model, a fact that is confirmed by the numerical analysis of the fully nonlinear model (see Sec.~\ref{length:sec}). 

Looking at $\mu_{\pm}^{(k_{\rm max}^{\pm})}$ we find that when the instability sets on, the eigenvalues are both fully real and negative. This implies that, in the long-time limit, the most unstable mode increases changing sign at every period, consistently with the presence of period doubling. (The period doubling fully develops after a transient, when nonlinear effects set in, and balance the unbounded exponential growth).

This information allows us to provide an analytical approximation for the boundary line. We have the most unstable mode for $k=\pi$ and it appears when both the eigenvalues become real. So, we can substitute $k=\pi$ inside Eq.~\eqref{eqig:eqn} and impose that the argument of the root is vanishing. In this way we find two solutions, one of them is
\begin{equation}\label{est:eqn}
  K^* = - 4 J + 2\sqrt{\gamma}+\gamma + 1\,.
\end{equation} 
We plot it in Fig.~\ref{patmax:fig}(b) and we see that it provides a very good approximation to the instability boundary line.

 On the opposite, $\Lambda_{+}$ -- the maximum of $|\mu_{+}^{(k)}|$ -- is independent of $J$, as we can see in Fig.~\ref{patmax:fig}(a). In the interval where $\Lambda_{+}$ is constant, the $|\mu_{+}^{(k)}|$ are independent of $k$, otherwise we have checked that the maximum occurs for $k_{\rm max}^{-}=0$. This happens for instance near $K=0$. In this region, substituting $k=0$ in Eq.~\eqref{eqig:eqn}, we can analytically check that  $\Lambda_{+}=1$ for $K=0$, whenever $0<\gamma<1$ (see also Fig.~\ref{patmax:fig}). For $K=0$, therefore, there is marginally stable mode with $|\lambda_{+}^{(0)}|=1$, that is fragile to the tiniest perturbation (see for instance~\cite{Berry_regirr78:proceeding}). It is therefore not surprising that in the limit $J\gg K$ the system can only show chaos (see the phase diagram in Fig.~\ref{phd:fig}).

\section{Pattern formation}\label{pt:sec}
\subsection{Examples of persistent patterns}
When the system is in the regime ``Trivial'' in Fig.~\ref{phd:fig}, all the values of $p_j$ relax to 0. Otherwise, in the regimes ``Pattern'' and ``Spatiotemporal order'', after a transient is died away, the system spontaneously forms patterns of $p_j$ that are persistent in time. We find that different initial conditions give rise to different asymptotic patterns, a phenomenon common in nonlinear dynamics.

\begin{figure}
  \begin{tabular}{c}
    \includegraphics[width=80mm]{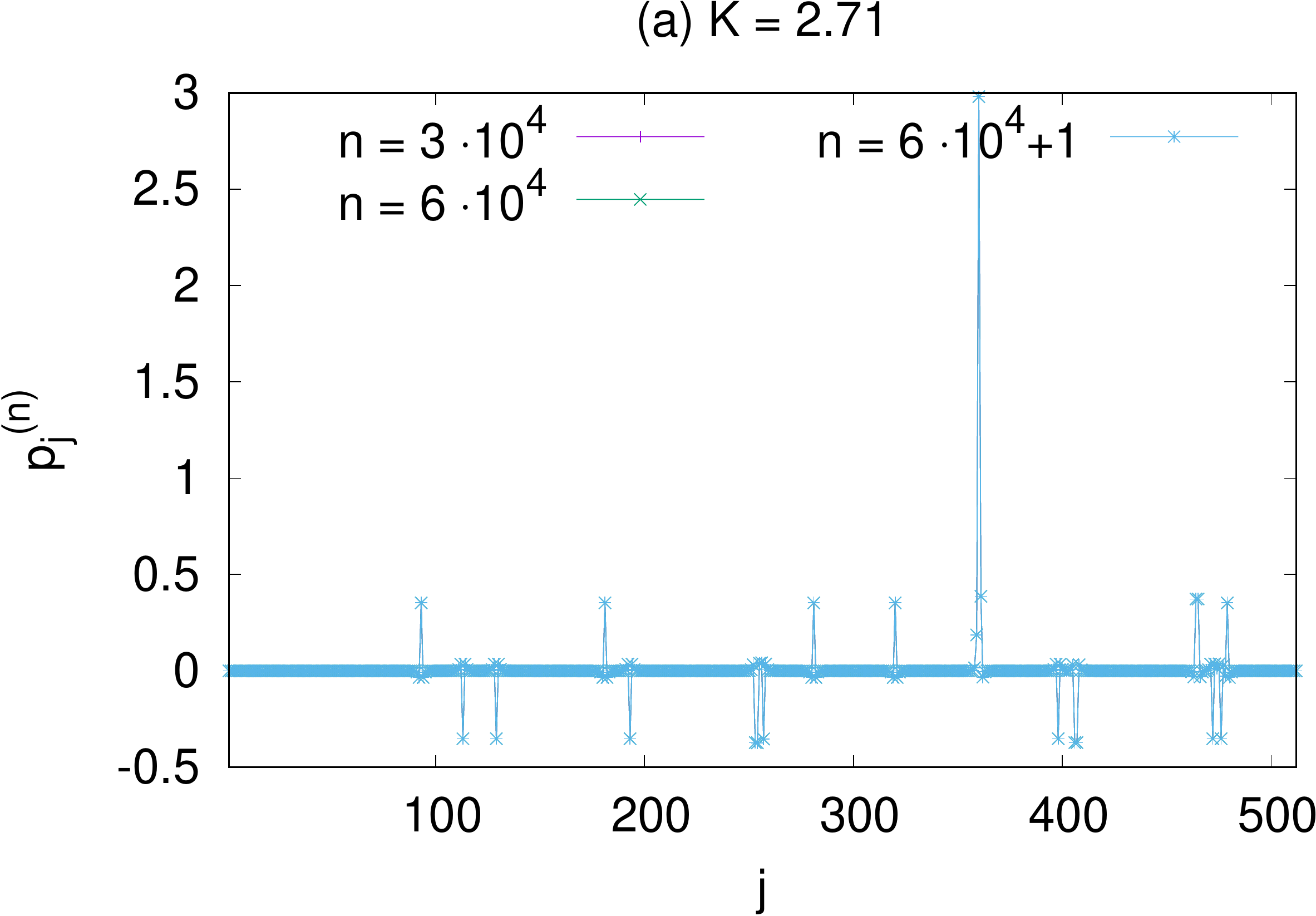}\\
    \includegraphics[width=80mm]{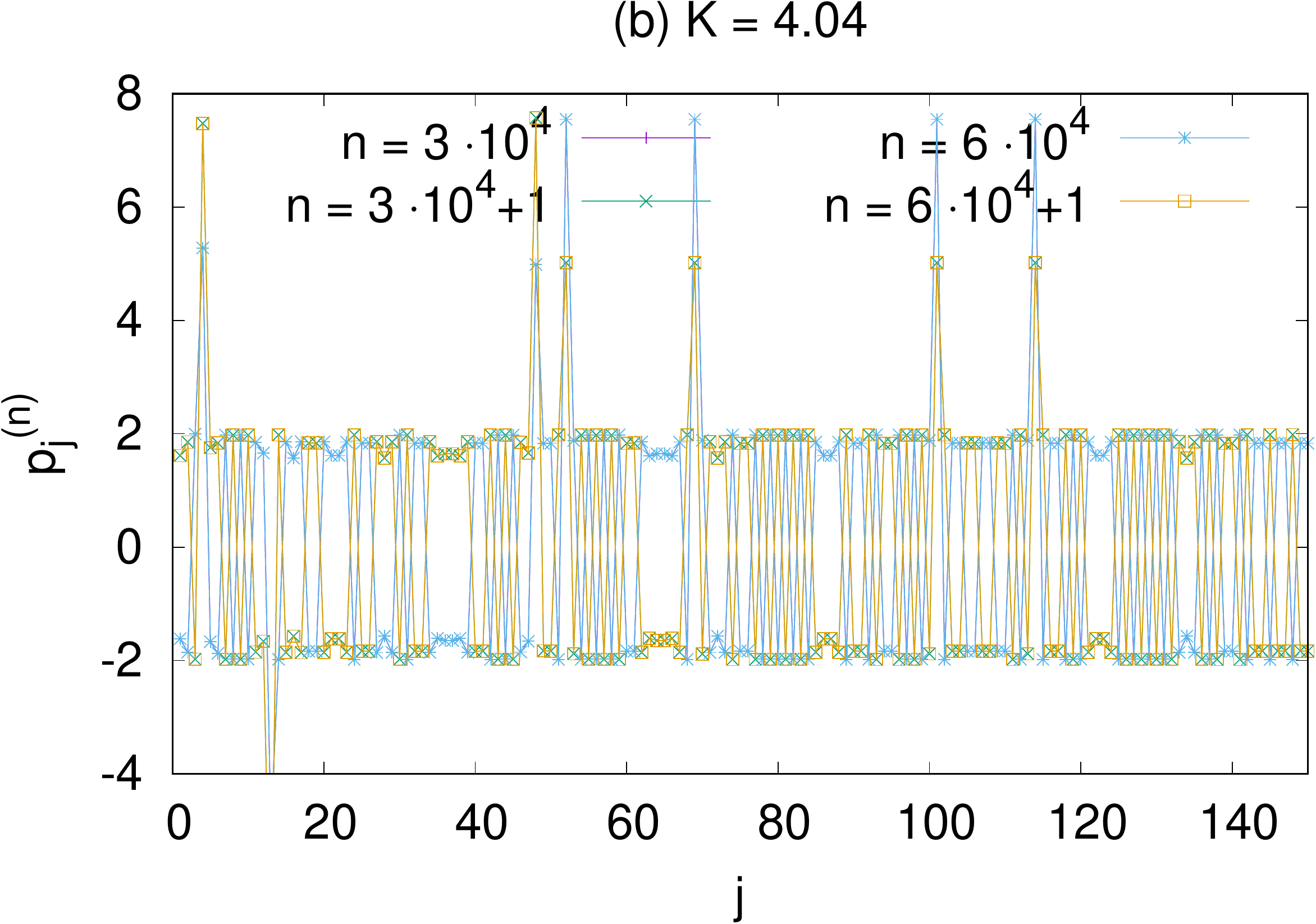}
  \end{tabular}
  \caption{Examples of asymptotic patterns. (a) Example for the regime ``Pattern'' in Fig.~\ref{phd:fig}. Here the pattern is persistent and independent of the stroboscopic time $n$ (Numerical parameters $J=0.1,\,K=2.71,\,L=512$, one given random initial condition). (b) Example for the regime ``Spatiotemporal ordering'' in Fig.~\ref{phd:fig}. Here the pattern is persistent and has a periodicity double than the driving, so the patterns for $n$ even coincide with each other, and the same with the patterns for $n$ odd. {Notice the regions with space periodicity double than the lattice separated by defects.} (Numerical parameters $J=0.1,\,K=4.04,\,L=512$, one given random initial condition. For clarity we have plotted only part of the pattern.)}\label{pat:fig}
\end{figure}

 In absence of period doubling (the regime ``Pattern'') the patterns are independent of the stroboscopic time. We show an example thereof in Fig.~\ref{pat:fig}(a). We initialize with one random initial condition and wait that the initial transient dies out. We see that the pattern is constant in the stroboscopic time, so it lies unchanged whatever the value of $n$. {Because the pattern comes back to itself after each cycle, the rotors behave in a perfectly synchronized way.}
 
 In the regime ``Spatiotemporal order'', instead, the patterns are associated with period doubling. In this regime, the persistence in time of the pattern means that it changes at each stroboscopic time, and comes back to itself after two cycles {(so also here the rotors are perfectly synchronized, with a period double than the driving)}. We show an example thereof in Fig.~\ref{pat:fig}(b). The period doubling appears in the fact that the pattern has a constant form for $n$ even ($n=3\cdot 10^4$, $n=6\cdot 10^4$) and a different equally constant form for $n$ odd ($n=3\cdot 10^4+1$, $n=6\cdot 10^4+1$).
 
  {Notice the staggered spatial ordering, with oscillations at a period double than the one of the lattice, separated by defects. This suggests that the most unstable mode with momentum $k=\pi$ provided by linear-stability analysis (see Sec.~\ref{stability:sec}) keeps its spatial periodicity double than the lattice also when it develops a stable pattern, as we are going to quantitatively confirm in the subsequent analysis.}
%
\subsection{Analysis of the pattern amplitude}
In order to study the existence and the properties of the patterns, it is important to quantify them. For that purpose we introduce two quantities. The first one is the pattern amplitude. To evaluate it, we fix $n\gg 1$, consider the variance of $p_j^{(n)}$ over space, average it over random initial state realizations, and evaluate the square root, namely
\begin{equation}
  \delta p_n = \left[\;\overline{\frac{1}{L}\sum_j(p_j^{(n)})^2-\left(\frac{1}{L}\sum_j p_j^{(n)}\right)^2}\;\right]^{1/2}\,.
\end{equation}

\begin{figure}[h!]
  \begin{tabular}{c}
    \includegraphics[width=80mm]{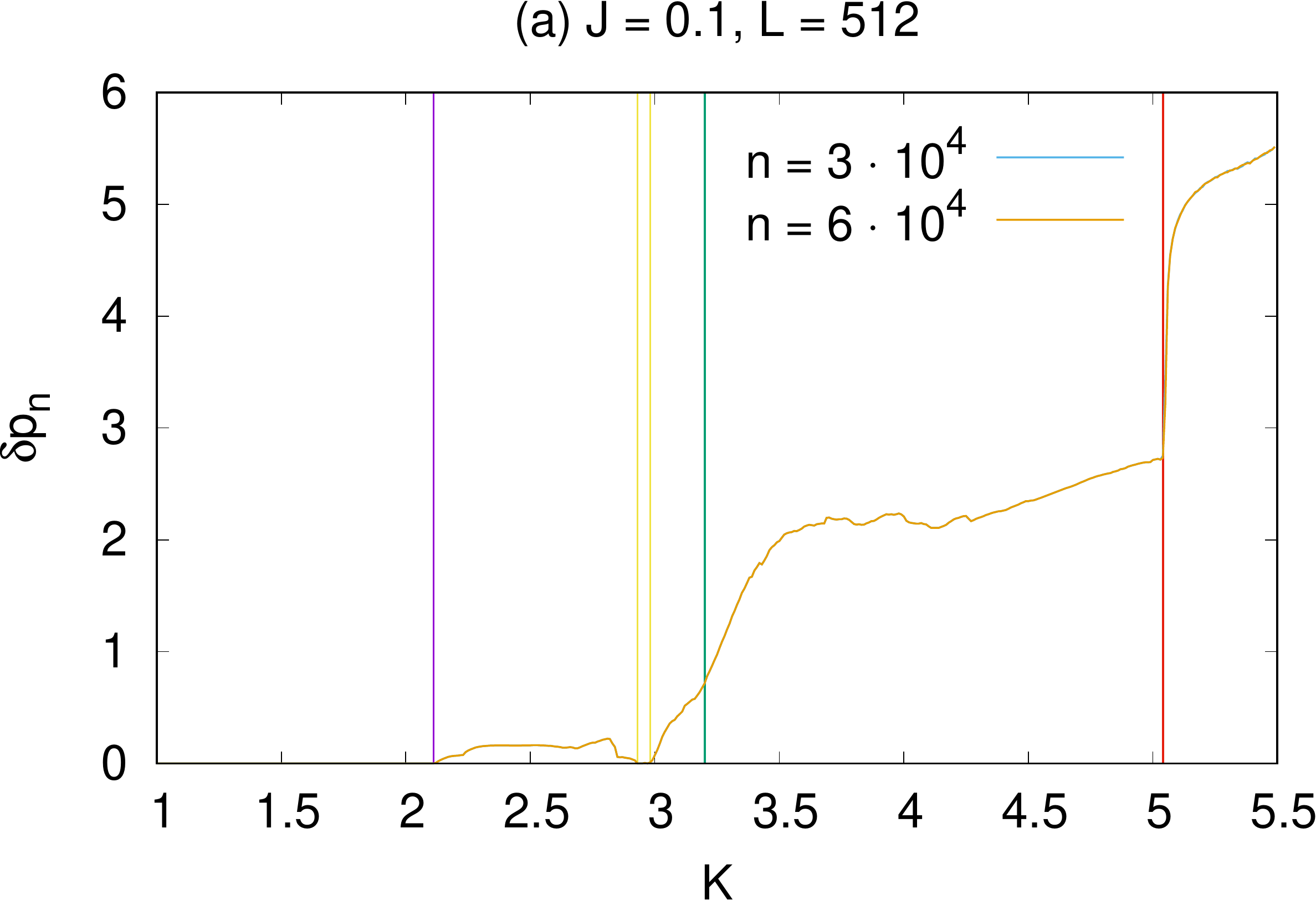}\\
    \includegraphics[width=80mm]{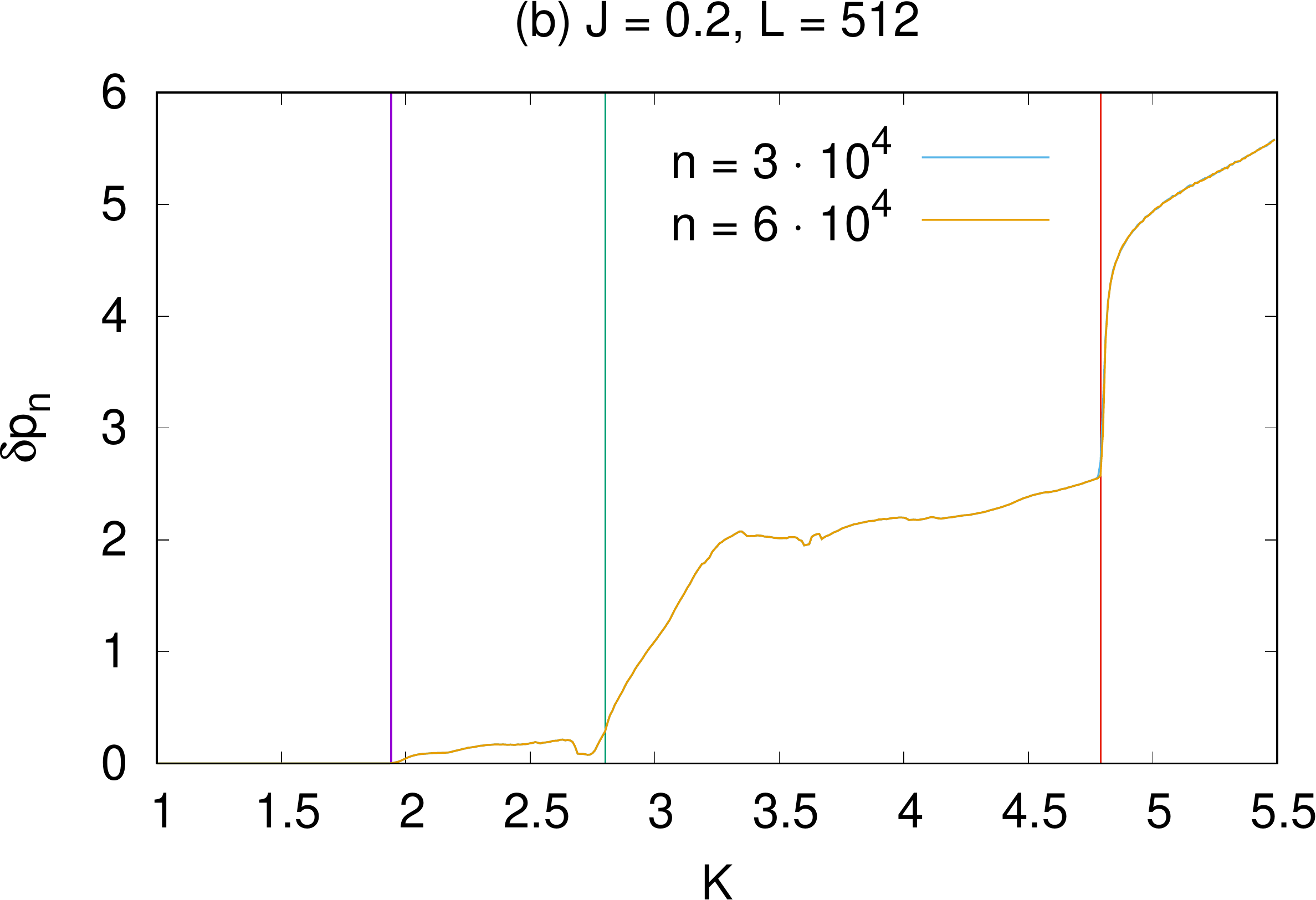}\\
    \includegraphics[width=80mm]{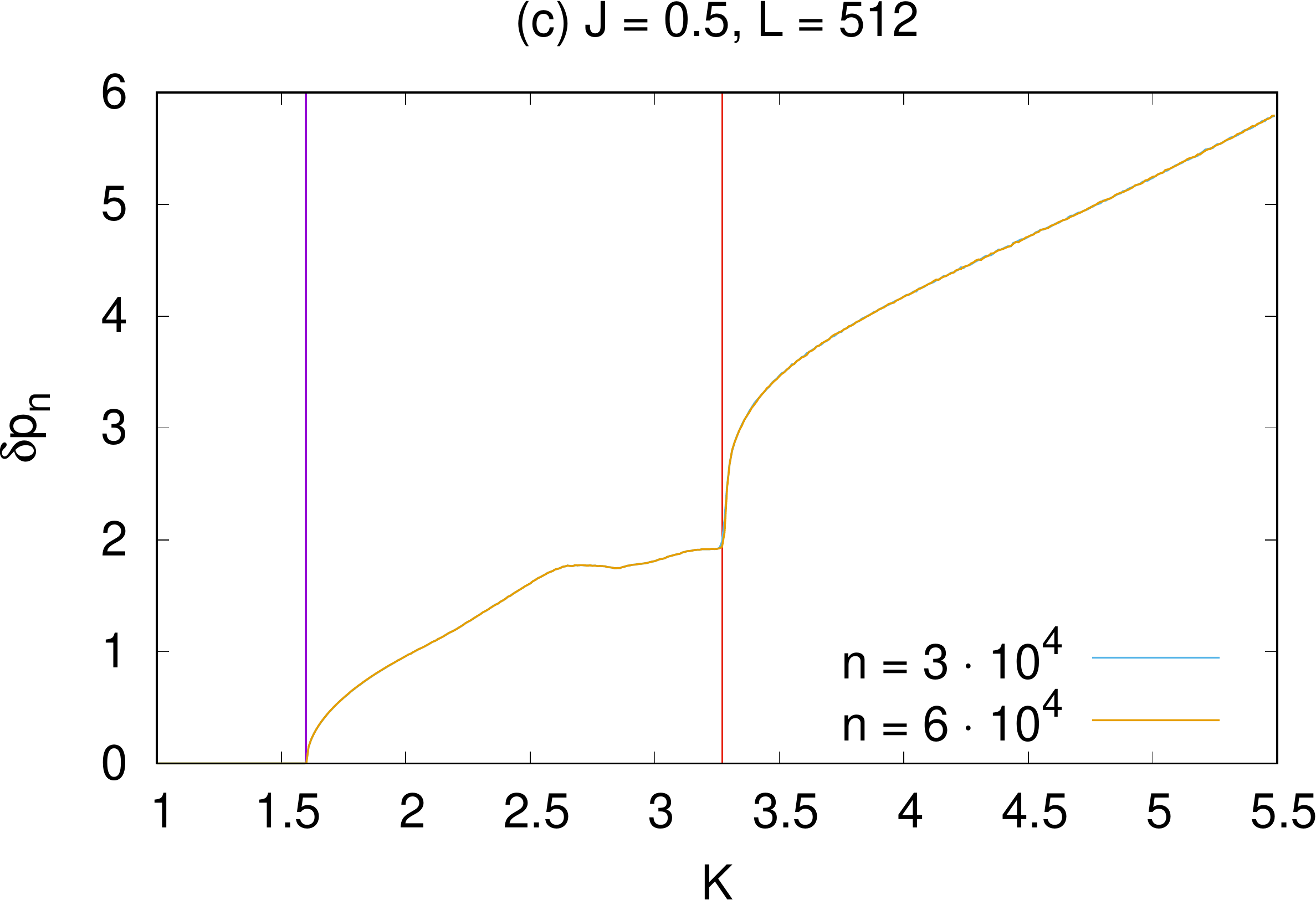}
  \end{tabular}
  \caption{$\delta p_n$ versus $K$ for different values of $J$ and of $n$ (chosen such that convergence has been attained). The vertical lines mark the boundaries of the different regimes listed in Fig.~\ref{phd:fig}: the red line bounds from below the chaotic regime, the green one the period-doubling (spatiotemporal ordering) regime, the purple one the patterning regime and between the yellow lines lies the weak patterning regime. The weak patterning regime exists only in panel (a), and in panel (c) the onset of patterning coincides with the onset of period doubling. Numerical parameters: $N_{\rm r} = 10^3,\,L=512$; (a) $J=0.1$, (b) $J=0.2$; (c) $J=0.5$.}\label{dpn:fig}
\end{figure}

This quantity is very important, because marks the existence of the patterns (it vanishes in the trivial state). We show some examples of $\delta p$ versus $K$ in Fig.~\ref{dpn:fig}. In all these figures we consider two values of $n$ ($n=3\cdot 10^4$ and $n=6\cdot 10^4$) to show that $\delta p_n$ has converged in time, and consider $L=512$, large enough that all finite-size effects have disappeared. Let us first focus our attention on the case $J=0.1$ [Fig.~\ref{dpn:fig}(a)]. We see first of all that $\delta p_n$ is independent of $n$ for $n\geq 3\cdot 10^4$. We see many features, let us discuss them moving from the right to the left. 

At the onset of chaos (red vertical line on the extreme right) we notice that $\delta p_n$ starts abruptly to increase with a discontinuous derivative. In the chaotic regime, patterns depend on time in an aperiodic fashion, but the average over initial-state realizations provides a $\delta p_n$ that does not depend on $n$. At the onset of the period doubling (green vertical line, second from the right) we see that $\delta p_n$ shows no special feature. 

For small $J$ the period-doubling regime is fully contained in a region where there is patterning ($\delta p_n> 0$) and the disappearance of period doubling gives no discontinuity, neither on $\delta p_n$ nor on its derivative. The situation is so for $J<0.45$; in contrast for $J\geq 0.45$ the threshold for patterning coincides with the one for period doubling. [We can see an example of that also in the plot in Fig.~\ref{dpn:fig}(c)]. It is important to emphasize that period doubling appears always in association with the appearance of patterns, that's why we define ``spatiotemporally ordered'' the range of parameters where period doubling appears.

\begin{figure}
  \begin{tabular}{c}
     \includegraphics[width=80mm]{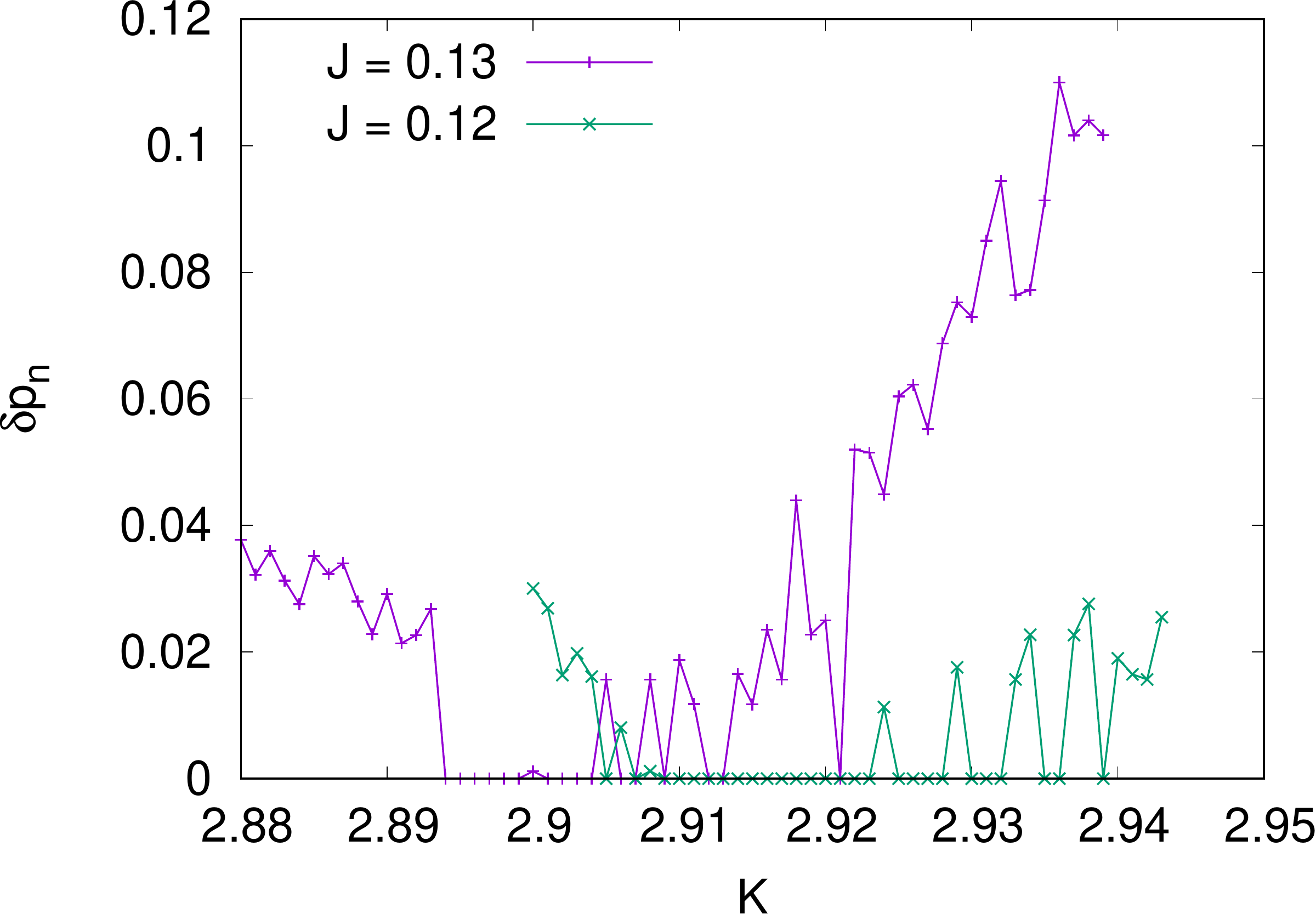}
  \end{tabular}
  \caption{Examples of $\delta p_n$ versus $K$ in the weak patterning regime. Numerical parameters: $L=160$, $N_{\rm r} = 10^3$.}\label{jagged:fig}
\end{figure}

Between the two yellow vertical lines in Fig.~\ref{dpn:fig}(a) there is the weak patterning regime. It is characterized by a jagged profile where very small values of $\delta p_n$ (order $10^{-3}$) alternate with vanishing values (and then no pattern at all). We show some magnification of this jagged profile in Fig.~\ref{jagged:fig}. This regime disappears for $J\simeq 0.14$, but also for larger $J$ we can see a marked local minimum of $\delta p_n$ for $J$ just below the onset of period doubling [see Fig.~\ref{dpn:fig}(b)].

For $J=0.5$ we are well inside the regime where the onset of patterning and of period doubling coincide, and patterns in the regular regime show always a time dependence of period 2 (spatiotemporal order) [Fig.~\ref{dpn:fig}(c)]. At the transition to chaos one can see the same features as in the other two cases.

\subsection{Analysis of the typical length scale of the pattern}\label{length:sec}
Another important property is the shape of the pattern. We can see in Fig.~\ref{pat:fig} that the patterns oscillate in space, and to these oscillations we can associate a wavelength, providing us the typical length scale of the pattern. 
We can estimate this length scale in the following way. Exploiting that the number of space oscillations is independent of $n$ (we can see some examples thereof in Fig.~\ref{pat:fig}), we choose $n\gg 1$ and numerically perform the space Fourier transform of $p_j^{(n)}$ as
\begin{equation}
  f_n(k) = \frac{1}{L}\sum_{j=1}^L\nep^{ik j} p_j^{(n)}\,,
\end{equation}
where $k=2\pi \ell /L$, with $\ell \in \{1,\,\ldots,\,L\}$ integer. Then we evaluate the power spectrum $|f_n(k)|^2$ and choose the value $k_{\rm max}^{(n)}$ where this power spectrum shows a maximum. The corresponding wavelength is $\lambda_{\rm max}^{(n)} = 2\pi/k_{\rm max}^{(n)}$. We perform the logarithmic average of this quantity over random realizations of the initial state and we get an estimate of the typical length scale of the patterns for a given set of parameters
\begin{equation}
  \lambda_n \equiv \exp\left(\overline{\log\lambda_{\rm max}^{(n)}}\right)\,.
\end{equation}
(We choose the logarithmic average because the distributions of the $\lambda_{\rm max}$ are broad).

\begin{figure}[h!]
  \begin{tabular}{c}
    \includegraphics[width=80mm]{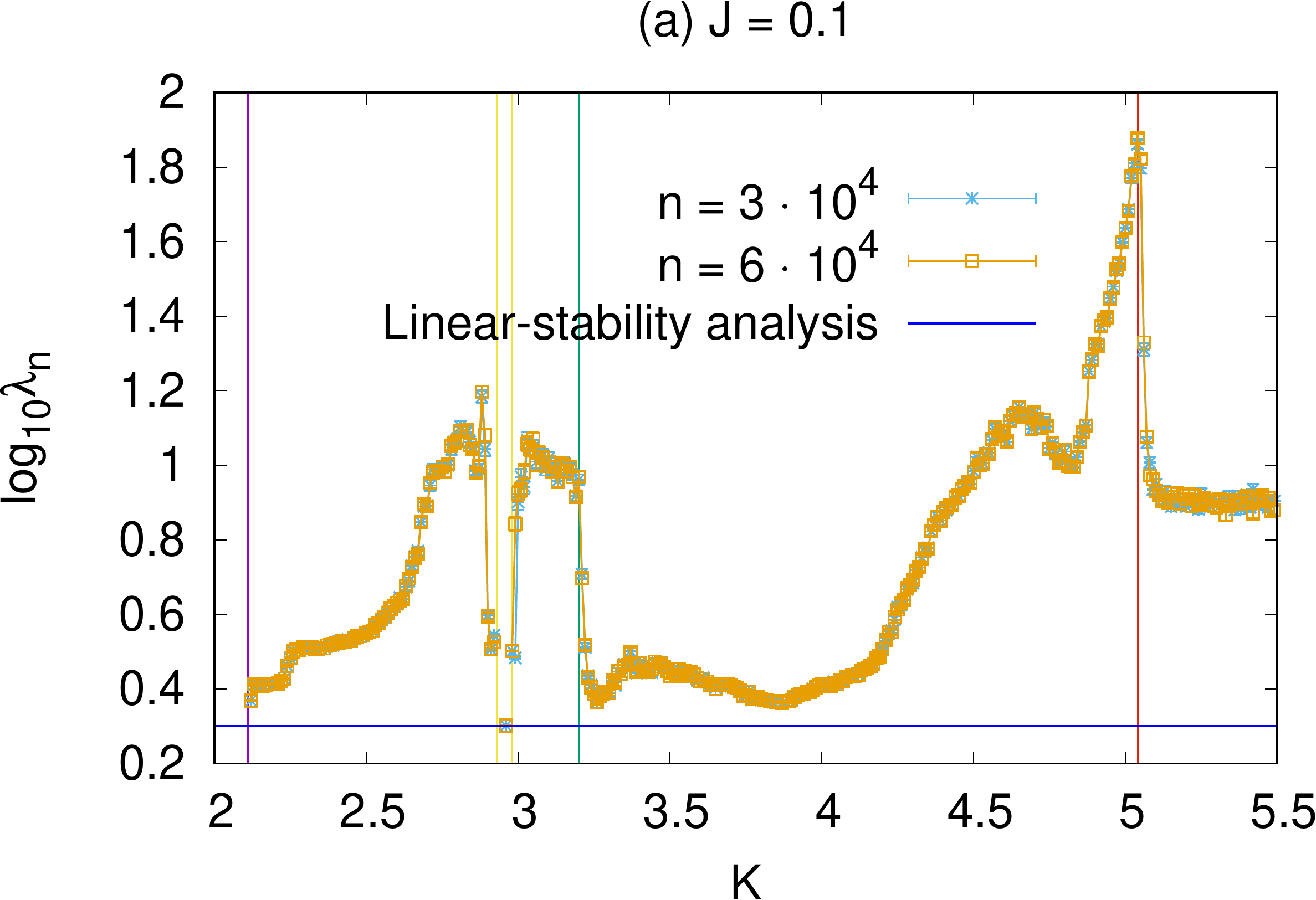}\\
    \includegraphics[width=80mm]{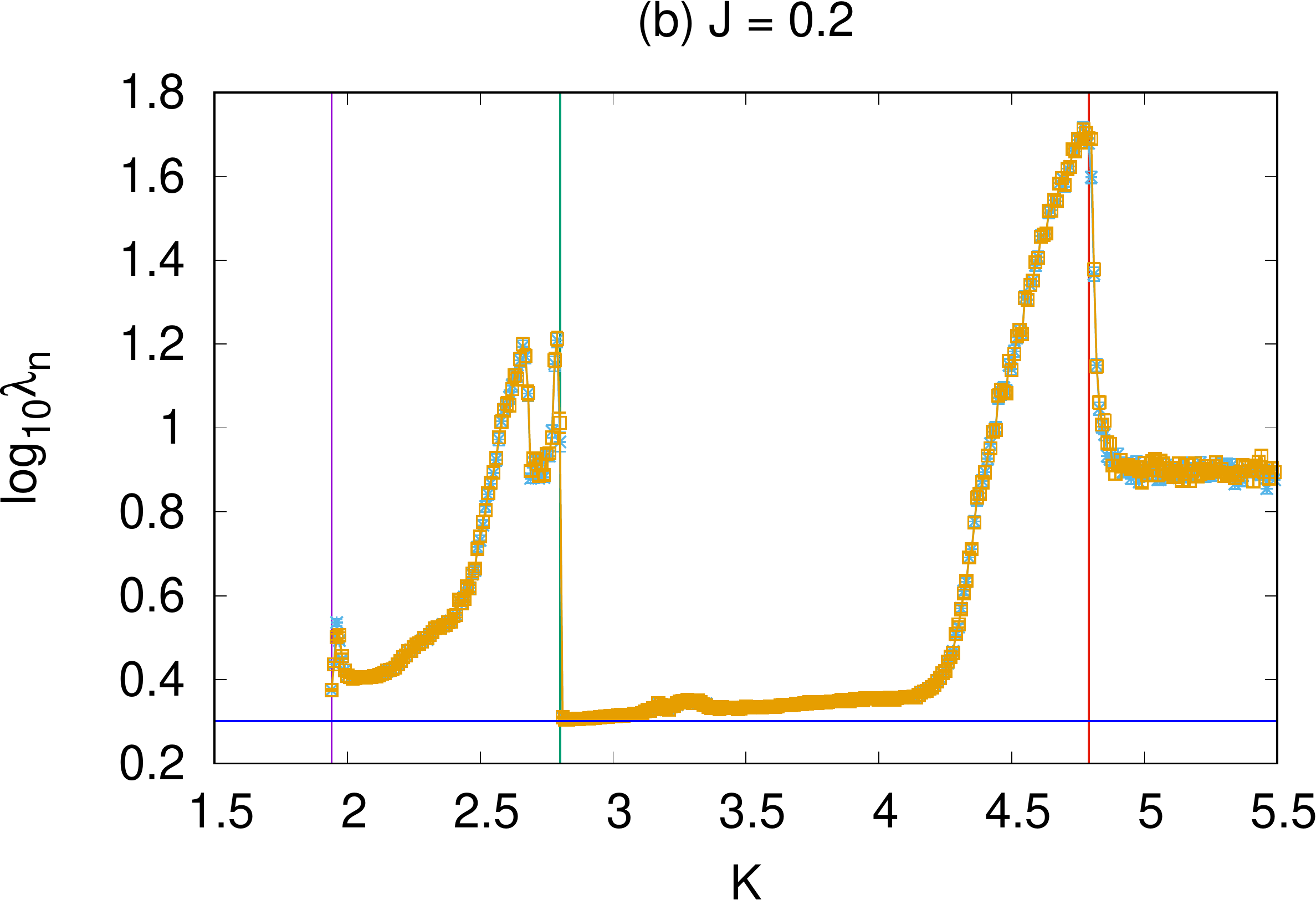}\\
    \includegraphics[width=80mm]{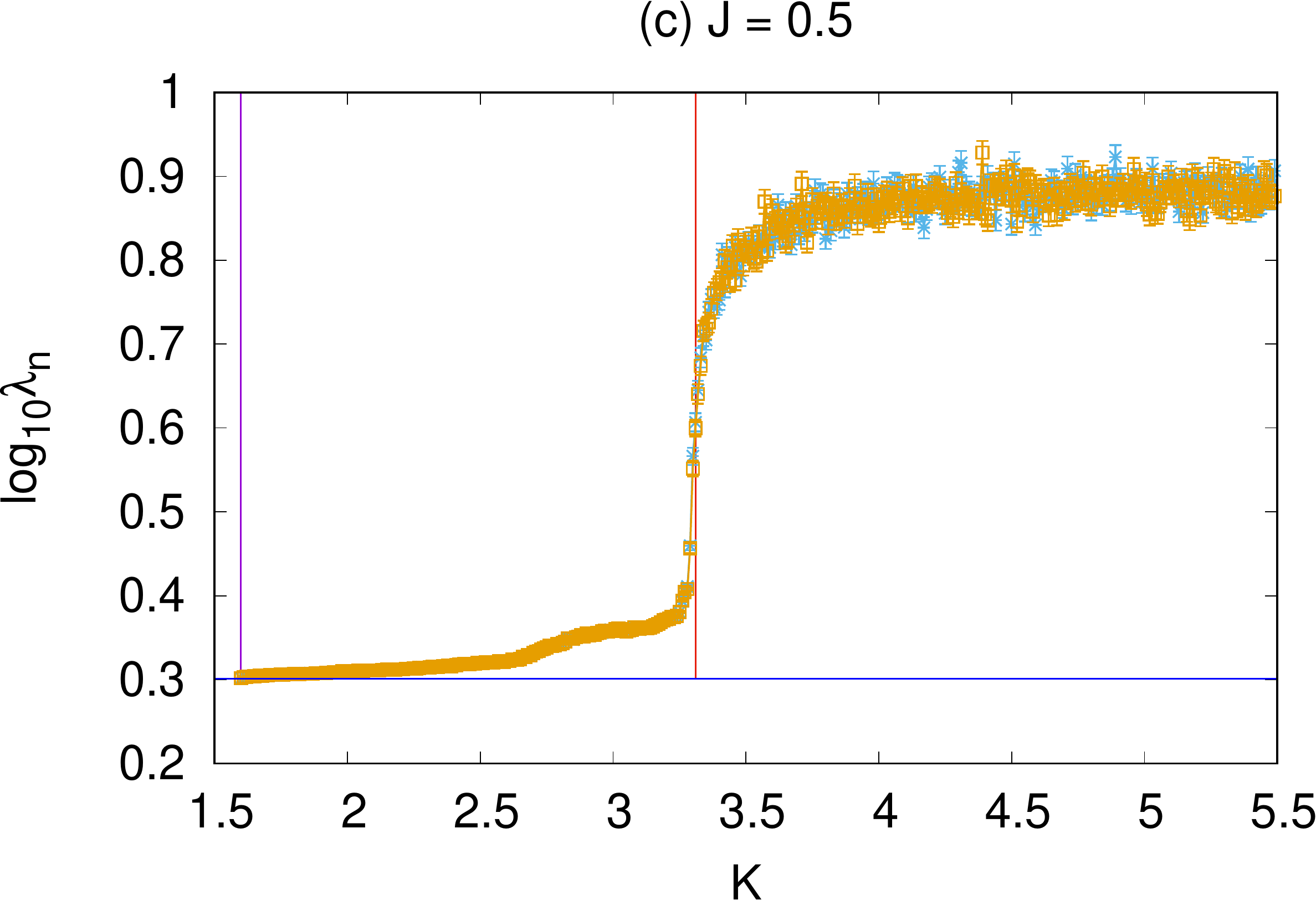}
%
  \end{tabular}
  \caption{Logarithm of the typical length scale of the patterns versus $\gamma$ for different system sizes and values of $J$ at time $n=10^5$. The errorbars are evaluated as $\delta\log_{10}\lambda^*=\frac{1}{\sqrt{N_{\rm r}}}\left[\overline{(\log_{10}\lambda)^2}-(\overline{\log_{10}\lambda})^2\right]^{1/2}$.  {The blue horizontal line marks the value $\lambda_n = 2$ predicted by the linear-stability analysis in presence of instability for the most unstable mode. There is a good agreement with $\lambda_n$ in the spatiotemporally ordered phase, near its lower phase boundary.} The parameters for each panel are written in the corresponding headings. $N_{\rm r}=10^3$.}\label{papat:fig}
\end{figure}

We plot some examples of $\log_{10}\lambda_n$ versus $K$, for $L=512$ and different values of $J$, in Fig.~\ref{papat:fig}. For $J=0.1$ [Fig.~\ref{papat:fig}(a)] we see a sudden drop inside the weak-patterning regime (in part of which $\lambda_n$ it is not even defined), and at the onset of the spatiotemporal ordered regime (dark green vertical line). So, although the amplitude of the patterns was not able to see the transition to the spatiotemporal ordering, it can be clearly seen in the typical length scale of the pattern. 

{The blue horizontal lines mark the value $\lambda_n = 2\pi/\pi = 2$ predicted by the linear-stability analysis for the most unstable mode. We find a good agreement with $\lambda_n$ inside the spatiotemporally ordered regime, when we consider $K$ near the lower phase boundary [the same occurs also for other values of $J$ -- see Fig.~\ref{papat:fig}(b,c)]. So, the linear-stability analysis predicts correctly not only the position of the lower phase boundary of the spatiotemporally ordered regime, but also the typical length scale of the pattern provided by the instability. }At the onset of chaos (red vertical line) the typical length scale shows ha huge peak, increasing of one order of magnitude and then suddenly dropping. In some sense at this transition the range of the correlations of the system increases.

For $J=0.2$ [Fig.~\ref{papat:fig}(b)] the weak patterning regime has disappeared, but $\log_{10}\lambda_n$ shows still a sudden drop at the onset of the spatiotemporal ordered regime and the peak at the onset of chaos.

For $J=0.5$ [Fig.~\ref{papat:fig}(c)] there are no time-independent patterns and the threshold for period doubling coincides with that of patterning, and there is no peak of the typical length scale at the onset of chaos.

\section{Chaos}\label{chaos:sec}
To quantify chaos, that's to say exponential increase in time of the distance of two nearby trajectories, we use the largest Lyapunov exponent (LLE -- see for instance~\cite{Piko,Ott}). It is defined in the following way. Considering two dynamics ${\bf X}^{(n)}=(p_1^{(n)},\,\ldots,\,p_L^{(n)},\,\theta_1^{(n)},\,\ldots,\,\theta_L^{(n)})$, ${{\bf X}^{(n)}}'=({p_1^{(n)}}',\,\ldots,\,{p_L^{(n)}}',\,{\theta_1^{(n)}}',\,\ldots,\,{\theta_L^{(n)}}')$ with different initial conditions ${\bf X}^{(0)}$, ${{\bf X}^{(0)}}'$ such that $||{\bf X}^{(0)}-{{\bf X}^{(0)}}'||=\epsilon>0$, the LLE is defined as
\begin{equation}
  \text{LLE} = \lim_{\epsilon\to 0}\lim_{n\to\infty}\frac{1}{n}\log\left(\frac{||{\bf X}^{(n)}-{{\bf X}^{(n)}}'||}{\epsilon}\right)\,,
\end{equation}
where $||\ldots||$ is the Euclidean norm. In a chaotic dynamics, this quantity evaluates the rate at which nearby trajectories exponentially separate from each other. So, when the dynamics is chaotic this quantity is positive, in absence of chaos it is vanishing or negative. To numerically evaluate the LLE we use the method explained in~\cite{PhysRevA.14.2338}.

The method goes as follows. One considers as above two initial conditions distant $\epsilon$. After the first cycle one evaluates the distance $d_1 = ||{\bf X}^{(1)}-{{\bf X}^{(1)}}'||$. Then one redefines ${{\bf X}^{(1)}}'$ as
\begin{equation} \label{redef:eqn}
  {{\bf X}^{(1)}}'' = {{\bf X}^{(1)}} + \frac{\epsilon}{d_1}({{\bf X}^{(1)}}'-{\bf X}^{(1)})\,,
\end{equation}
so that the distance between ${\bf X}^{(1)}$ and  ${{\bf X}^{(1)}}''$ becomes $\epsilon$ again. With these initial conditions one performs another stroboscopic-evolution step getting some ${\bf X}^{(2)}$ and  ${{\bf X}^{(2)}}'$. So one gets another value of the distance $d_2$, and performs a redefinition of ${{\bf X}^{(2)}}'$ as in Eq.~\eqref{redef:eqn}. This cycle is repeated many times, do that one gets a sequence of distances $d_1,\,d_2,\,d_3,\,\ldots,\,d_n$ and the Lyapunov exponent is given by
\begin{equation}\label{lyapel:eqn}
  \text{LLE}=\frac{1}{\mathcal{T}}\sum_{k=1}^{\mathcal{T}}\log\frac{d_k}{\epsilon}\,,
\end{equation}
where $\mathcal{T}$ is large enough and $\epsilon$ small enough so that convergence has been attained.

We use precisely this formula to get the LLE. We choose ${\bf X}^{(0)}$ taking all the $p_j^{(0)}$ and the $\theta_j^{(0)}$ from a random distribution uniform in the interval $[-1,1]$. We take ${{\bf X}^{(0)}}'$ equal to ${\bf X}^{(0)}$ everywhere but on the coordinate ${p_1^{(0)}}'$ that we take ${p_j^{(0)}}'=p_j^{(0)}+\epsilon$. To make convergence faster, we average Eq.~\eqref{lyapel:eqn} over $N_{\rm r}$ realizations of the random ${\bf X}^{(0)}$.

\begin{figure}
  \begin{tabular}{c}
      \includegraphics[width=80mm]{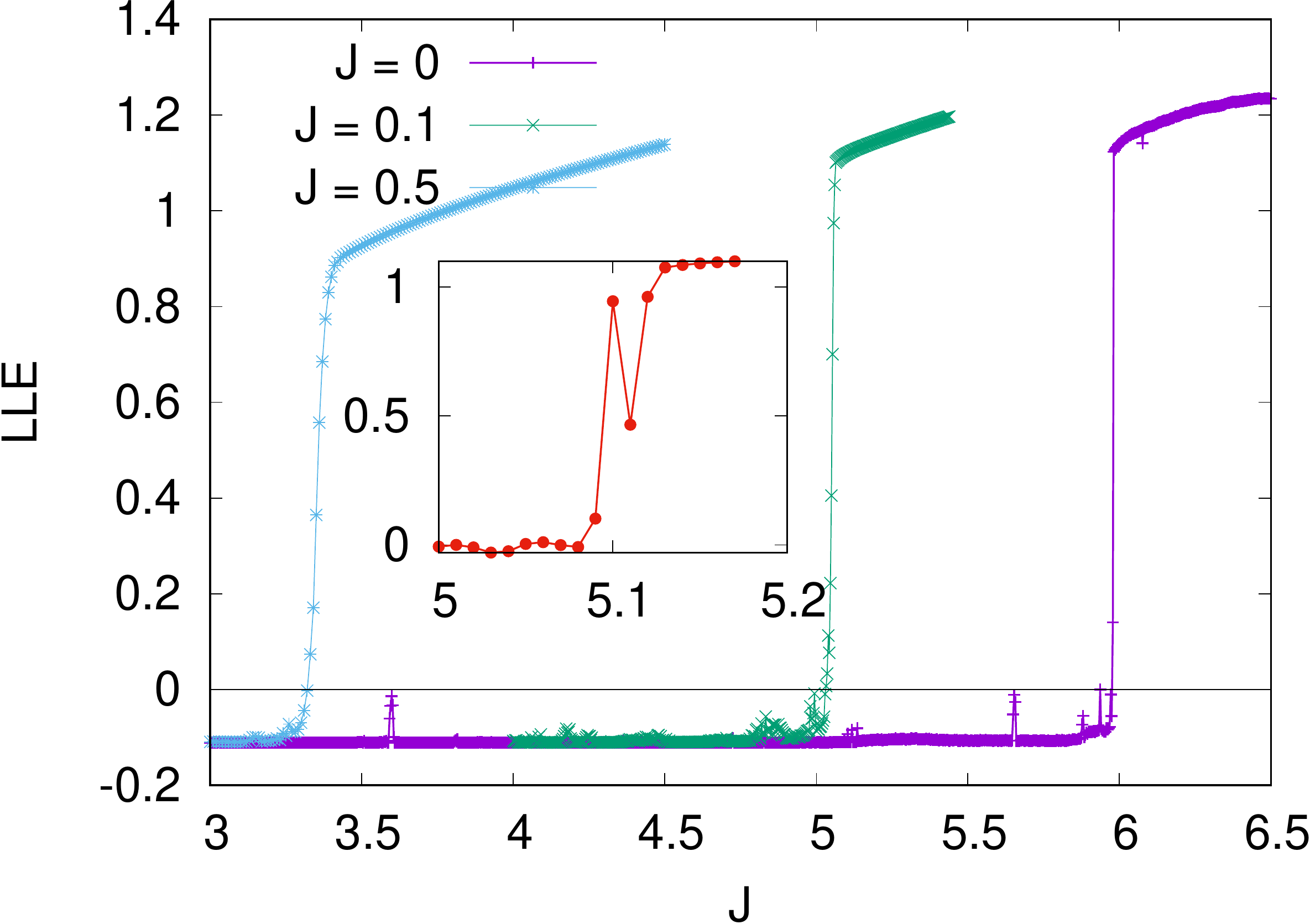}
    \end{tabular}
     \caption{LLE versus $K$ for different values of $J$ (In the inset $J=0.04$ is plotted). For each curve we choose the value of $\mathcal{T}$, $N_{\rm r}$ and $L$ so that convergence is attained (for $J=0$, $\mathcal{T}=10^5,\,N_{\rm r} = 100,\,L=1$; for $J=0.04$, $\mathcal{T}=3\cdot 10^5,\,N_{\rm r}=300,\,L=160$; for $J=0.1$, $\mathcal{T}=10^5,\,N_{\rm r}=100,\,L=40$; for $J=0.5$,  $\mathcal{T}=5\cdot 10^5,\,N_{\rm r}=500,\,L=40$.}\label{ly:fig}
\end{figure}

Fixing $J$, we find a quite sudden transition in $K$ from regular behavior ($\text{LLE}<0$) to chaotic behavior ($\text{LLE}>0$), provided the system size is large enough. This allows to map the boundary line of the chaotic region shown in Fig.~\ref{phd:fig}. The only region where the mapping of this line is problematic is around $J=0.04$. Here, there is not a sharp transition from a negative LLE to a positive one, but a range (the segment in red in Fig.~\ref{ly:fig}) where the Lyapunov exponent lies near 0 and often becomes slightly positive (see inset of Fig.~\ref{ly:fig}). This value of $J$ marks an abrupt change in the behavior of the boundary line of the chaotic region, that for $J<0.04$ keeps a value similar to the single-particle case ($J=0$) and for $J>0.04$ starts going down as a straight line (see Fig.~\ref{phd:fig}). 

{In the limit of $J\gg K$ there is only chaos, consistently with the numerical stability analysis described above, that for $K=0$ provides a marginally stable (and therefore very fragile) mode.}
\section{Period $m$-tupling}\label{mtup:sec}
\subsection{Bifurcation diagram of the single rotor}
\begin{figure}
  \begin{tabular}{c}
      \includegraphics[width=80mm]{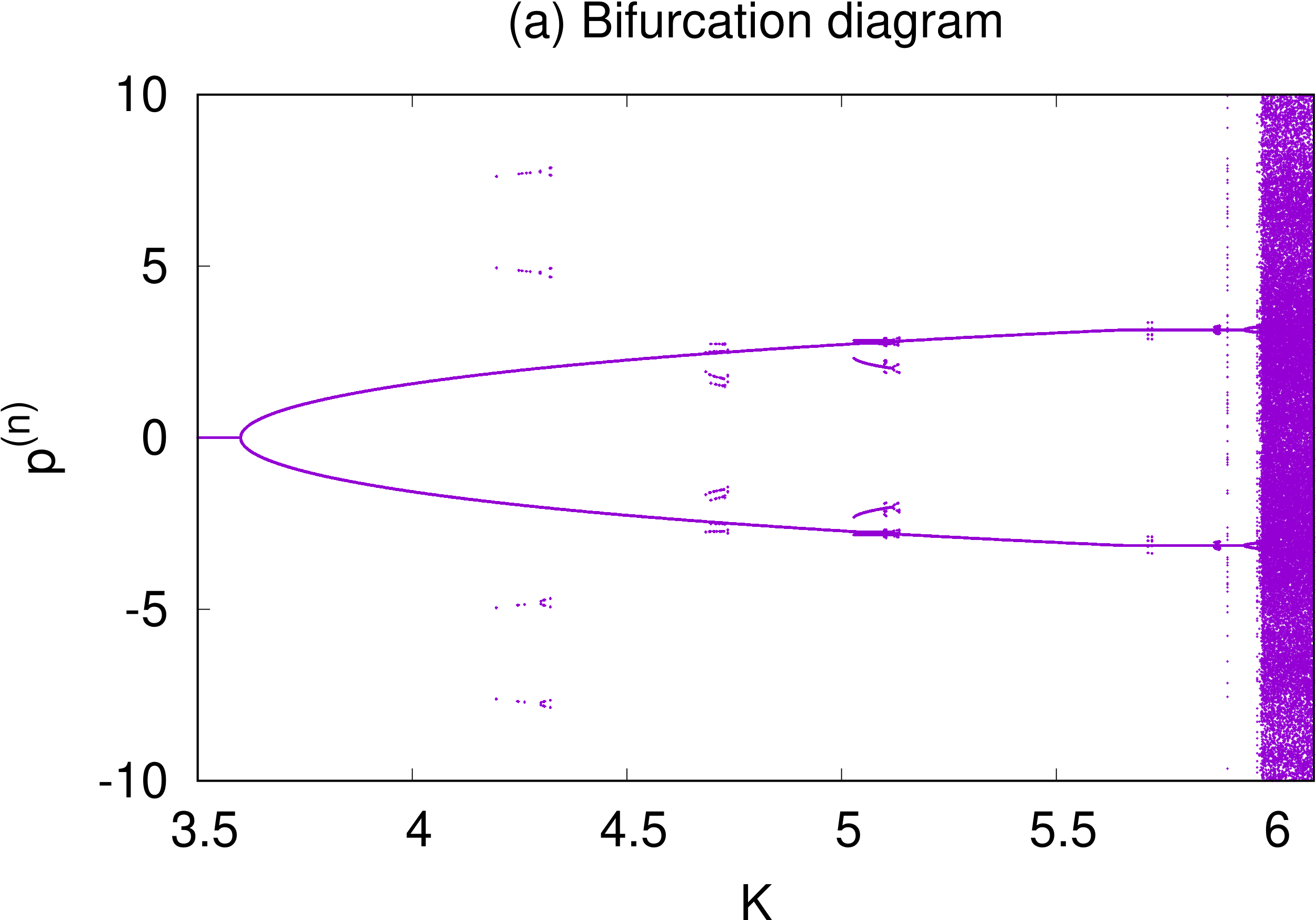}\\
      \includegraphics[width=80mm]{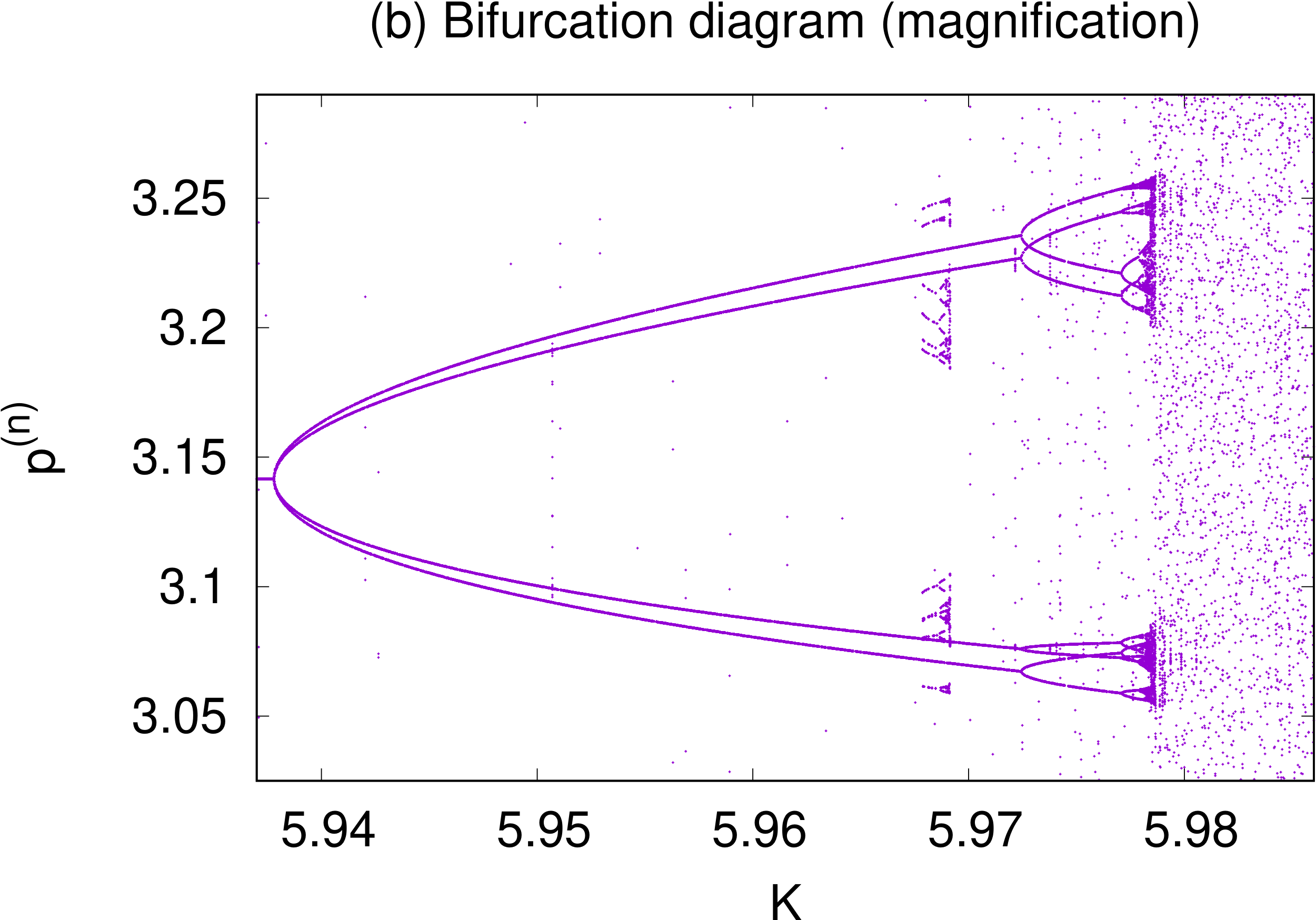}
    \end{tabular}
     \caption{(Panel a) Bifurcation diagram. (Panel b) A magnification thereof near the upper branch just below the onset of chaos. Notice the two main parallel bifurcation cascades (for each value of $K$ there are points on only one of the two).}\label{bd:fig}
\end{figure}

In order to discuss period $m$-tupling, let us start with the single dissipative rotor model Eq.~\eqref{sing:eqn}. This model displays a period-doubling cascade similar to the standard one seen in the logistic map. Before discussing it more quantitatively, it is useful to show it by means of a bifurcation diagram. In the plots of Fig.~\ref{bd:fig} we put $K$ on the horizontal coordinate, and -- for each value of $K$ -- we write on the vertical coordinate the last $10^3$ stroboscopic values of $p^{(n)}$, for an evolution lasting $\mathcal{T}=10^5$ periods. If for that $K$ the system relaxes to an asymptotic stroboscopic value, we see a single value on the vertical coordinate; if there is a period doubling we will see two values, if there period $m$-tupling for generic $m$ we will see $m$ values.

What we see is a period-doubling cascade, that's to say a sequence of points of $K$ where pitchfork bifurcations occur and the number of points doubles. (See~\cite{baba:book},~\cite{strogatz:book},~\cite{gold:book} for more details on pitchfork bifurcations and period-doubling cascade). So one moves to period doubling, to period 8-tupling, to period 16-pling (all the powers of 2). The period doubling cascade ends at the onset of chaos, that here starts at $K\simeq 5.978$ as one can see in the bifurcation plot as the appearance of a region with randomly scattered points (one can confirm this chaos threshold with the LLE analysis explained in Sec.~\ref{chaos:sec}). This period doubling cascade is similar to the logistic map, but there is more, because there are two parallel bifurcation cascades, as one can see in Fig.~\ref{bd:fig}(b), and beyond that also some small ones near $K=5.97$. Quite remarkably, for each value of $K$, the system chooses just one of the bifurcation cascades, in an apparently random way.
\subsection{Definition of the period $m$-tupling order parameters}
In order to better quantify these phenomena in a way that can be generalized to the many-body case, we take inspiration from the literature on discrete time crystals~\cite{Else_2016}, and define the following period $m$-tupling onsite order parameters
\begin{equation}
  \mathcal{O}_n^m(j)=\left[{p_j^{(n)}}\cos\left(\frac{2\pi n}{m}\right)\right]\,,
\end{equation}
then we average them over time and sites, and 
\begin{equation}
  \mathcal{O}^m = \frac{1}{n_0}\sum_{n=\mathcal{T}-n_0}^{\mathcal{T}}\frac{1}{L}\sum_{j=1}^L\mathcal{O}_n^m(j)\,,
\end{equation}
where we choose $n_0$ and $\mathcal{T}$ so that the initial transient is vanished and the sum over $n$ is converged. Our order parameter is the average over random initial-state realizations [this average is represented by the symbol $\overline{(\ldots)}$] of the absolute value of this quantity
\begin{equation}\label{om:eqn}
  \mathcal{O}^{(m)} = \overline{|\mathcal{O}^m|}\,.
\end{equation}
%
It is not difficult to convince oneself, that -- for $m>1$ -- if $\mathcal{O}_n^m$ shows a response at frequency $m$, then the average over an infinite time of $\mathcal{O}_n^m$ is nonvanishing. Notice that the system could show the linear superposition of responses with different periods, so a nonvanishing $\mathcal{O}_n^m$ is a necessary condition for a period $m$-tupling, but not sufficient. For instance, there can be also a response with period $2m$ -- thereby a period $2m$-tupling -- and we could still find a nonvanishing $\mathcal{O}_n^m$, as we are going to see below.

It is important to stress that these order parameters are of some usefulness only in the regular regime where $\text{LLE}<0$. Where there is chaos ($\text{LLE}>0$) the dynamics of the $p_j^{(n)}$ shows aperiodic oscillations, there is a response at all the frequencies (see for instance~\cite{pomeau:book}), and so all the $\mathcal{O}^{(m)}$, providing no information on the existence of a period doubling, but just trivially witnessing the chaos in the dynamics. So we will focus our period $m$-tupling analysis on the regime of regular dynamics.

We define also the quadratic average
\begin{equation}\label{o2m:eqn}
  \mathcal{O}^{(2,m)} =\overline{\left[\frac{1}{n_0}\sum_{n=\mathcal{T}-n_0}^{\mathcal{T}} \frac{1}{L}\sum_{j=1}^L\mathcal{O}_n^m(j)\right]^2}\,,
\end{equation}
and obtain the uncertainty on $\mathcal{O}^{(m)}$ as
\begin{equation}\label{fluc:eqn}
  \delta \mathcal{O}^{(m)} = \frac{1}{\sqrt{N_r}}\sqrt{ \mathcal{O}^{(2,m)}-(\mathcal{O}^{(m)})^2}\,,
\end{equation}
where $N_r$ is the total number of randomness/noise realizations. In all the numerical analyses that follow, we choose $\mathcal{T}$ finite large enough so that the limits in Eqs.~\eqref{om:eqn} and~\eqref{o2m:eqn} are converged.
\subsection{Results of the order-parameter analysis}\label{double:sec}
\begin{figure}
  \begin{tabular}{c}
       (a)\\
      \includegraphics[width=80mm]{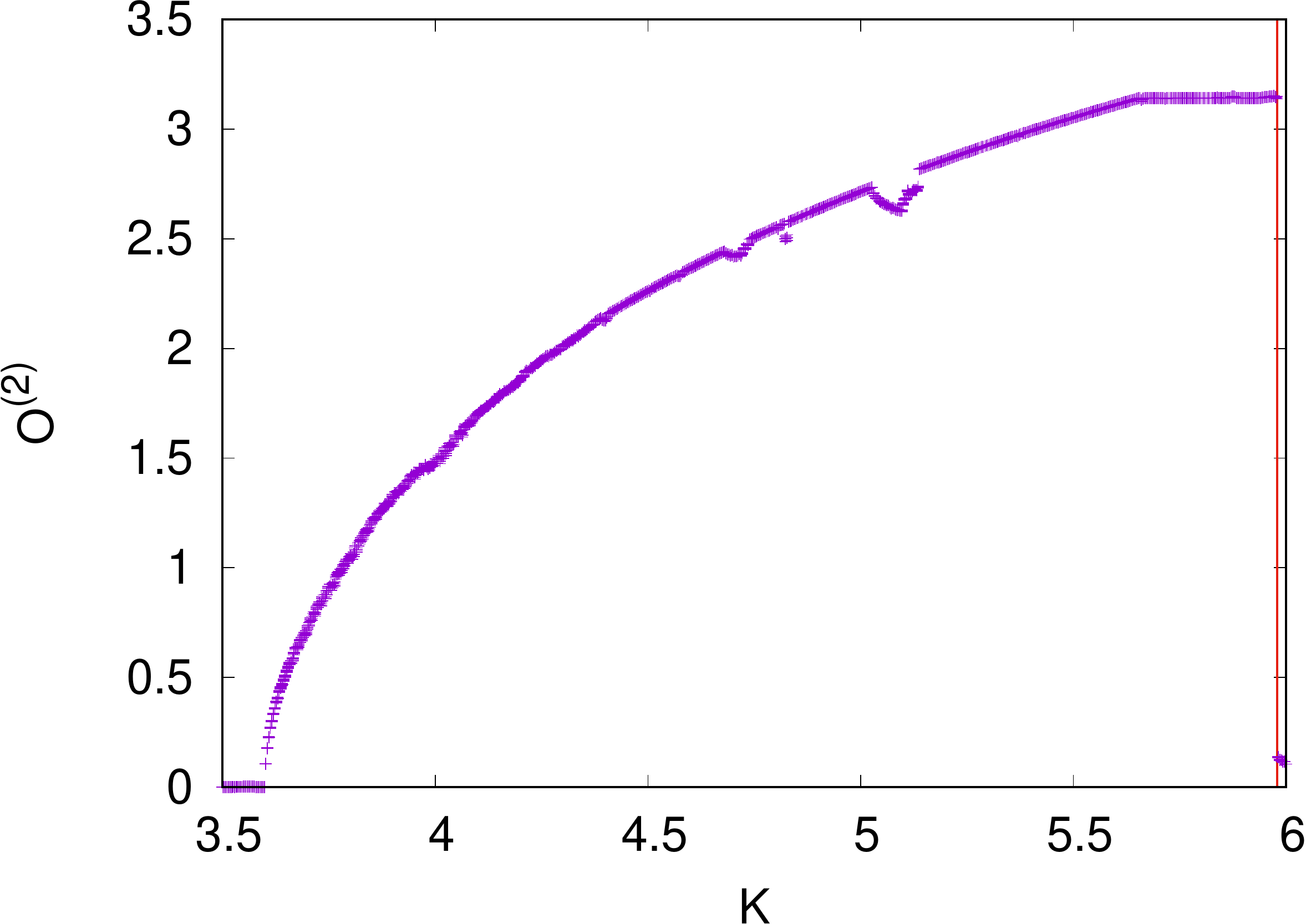}\\
      (b)\\
      \includegraphics[width=80mm]{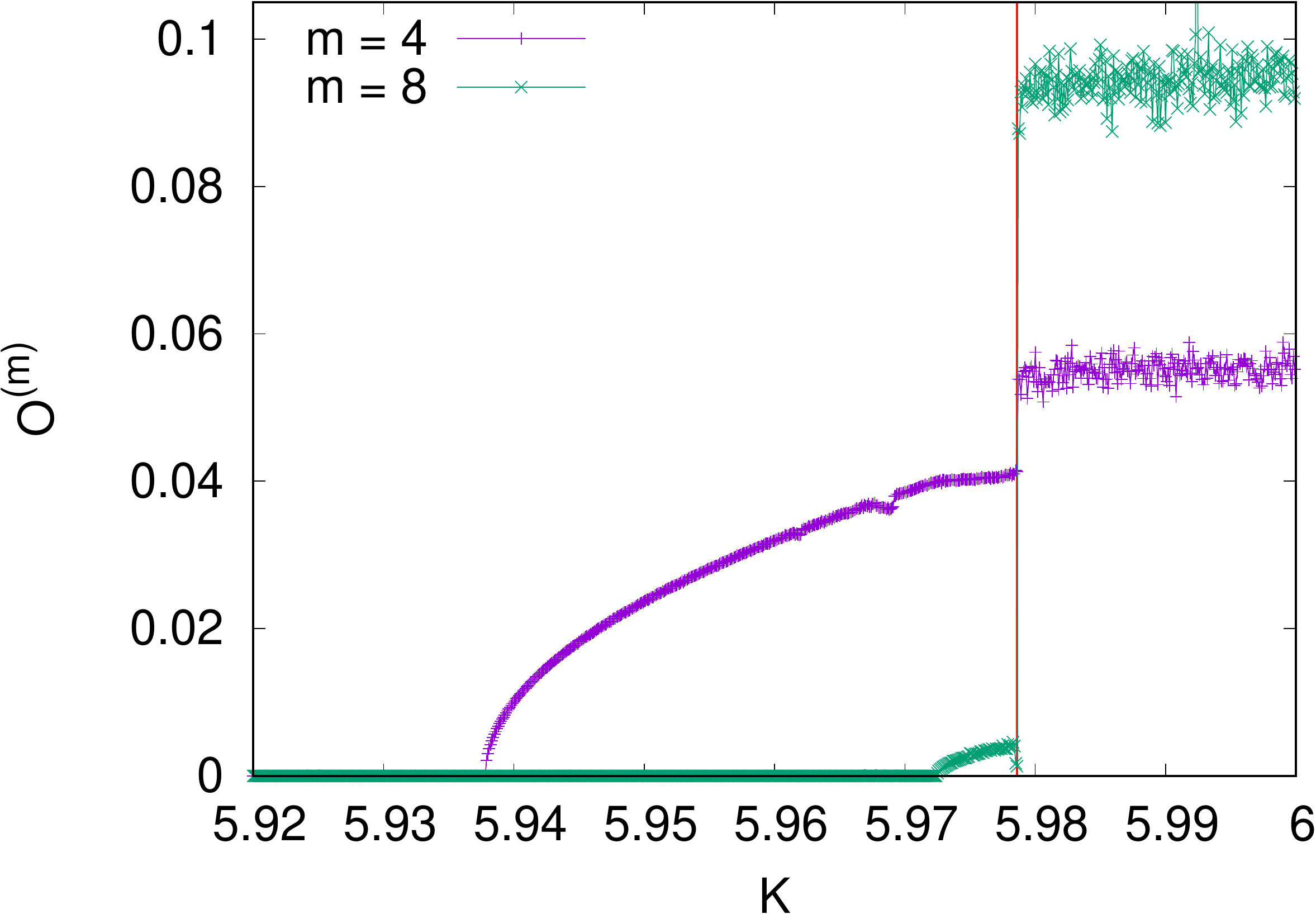}
    \end{tabular}
     \caption{Single rotor. (Panel a) Period-doubling order parameters versus $K$. (Panel b) Period $4$-tupling and period $8$-pling order parameter versus $K$. The vertical line marks the onset of chaos, as found with the LLE analysis. Numerical parameters $\mathcal{T}=4\cdot 10^4,\,n_0=10^3,\, N_{\rm r}=10^3$.}\label{op:fig}
\end{figure}

Let us start with the case of the single rotor. We show $\mathcal{O}^{(m)}$ versus $K$ for $m=2,\,4,\,8$ in Fig.~\ref{op:fig}. We see that the behavior of the order parameters closely mirrors the one of the bifurcation diagram in Fig.~\ref{bd:fig}. $\mathcal{O}^{(2)}$ is nonvanishing whenever there is at least period doubling [Fig.~\ref{bd:fig}(a)], $\mathcal{O}^{(4)}$ is nonvanishing whenever there is at least period 4-tupling, and $\mathcal{O}^{(8)}$ is nonvanishing whenever there is at least period 8-tupling [Fig.~\ref{bd:fig}(b)]. We see also that the response at $m=4$ is two orders of magnitude smaller than the one at $m=2$, and the one at $m=8$ even smaller. Also this property closely mirrors the fact that at each bifurcation the outcoming branches are much nearer than the ones at the bifurcation before (see Fig.~\ref{bd:fig} and the self-similarity analysis in~\cite{Feigenbaum}).

Of course the analysis is meaningful whenever the dynamics is regular. After the onset of chaos (marked by the vertical line in Fig.~\ref{bd:fig}) all the order parameters are nonvanishing and of the same order of magnitude, because the chaotic dynamics has contributions at all the frequencies.

\begin{figure}
  \begin{tabular}{c}
       (a)\\
      \includegraphics[width=80mm]{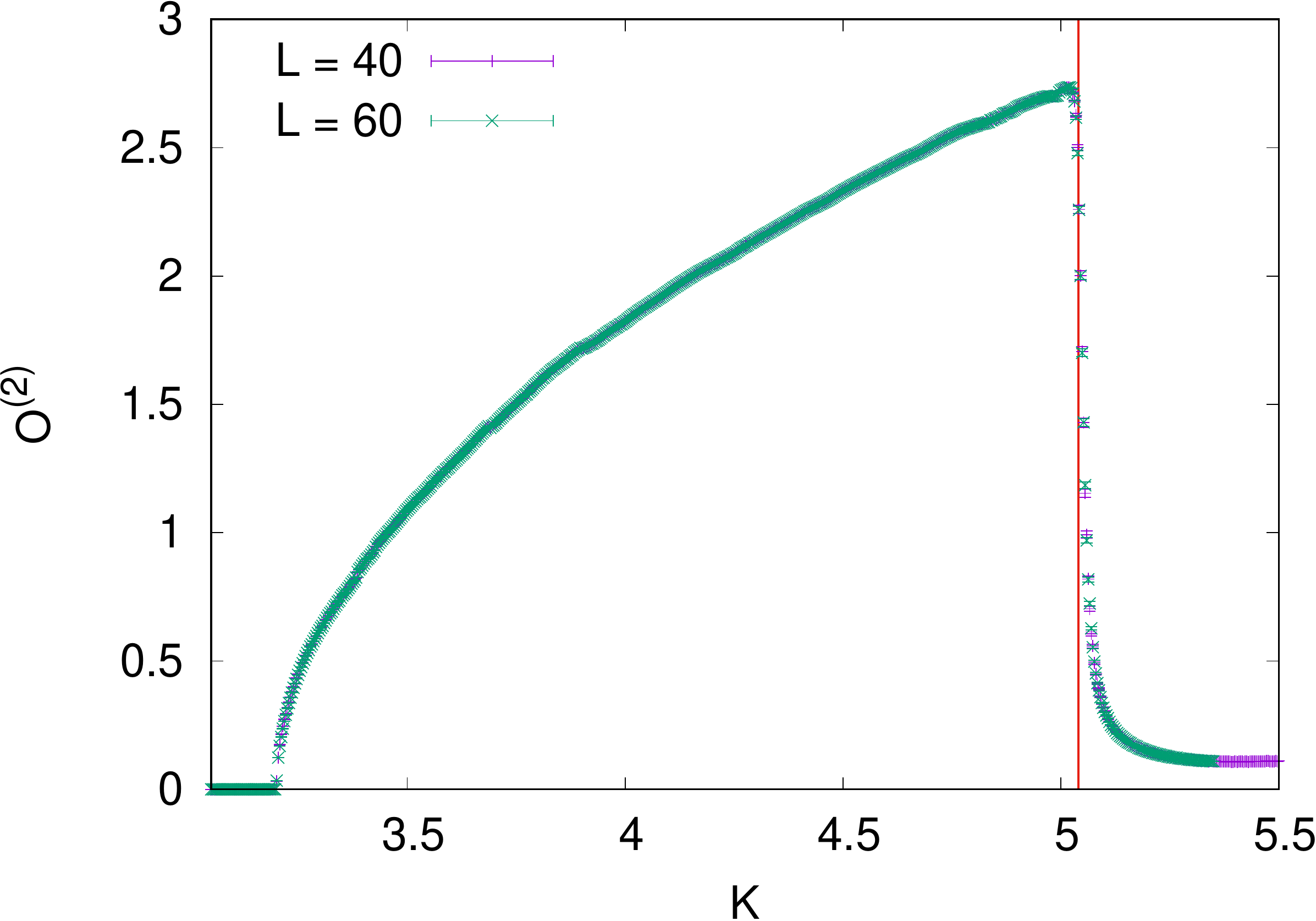}\\
      (b)\\
      \includegraphics[width=80mm]{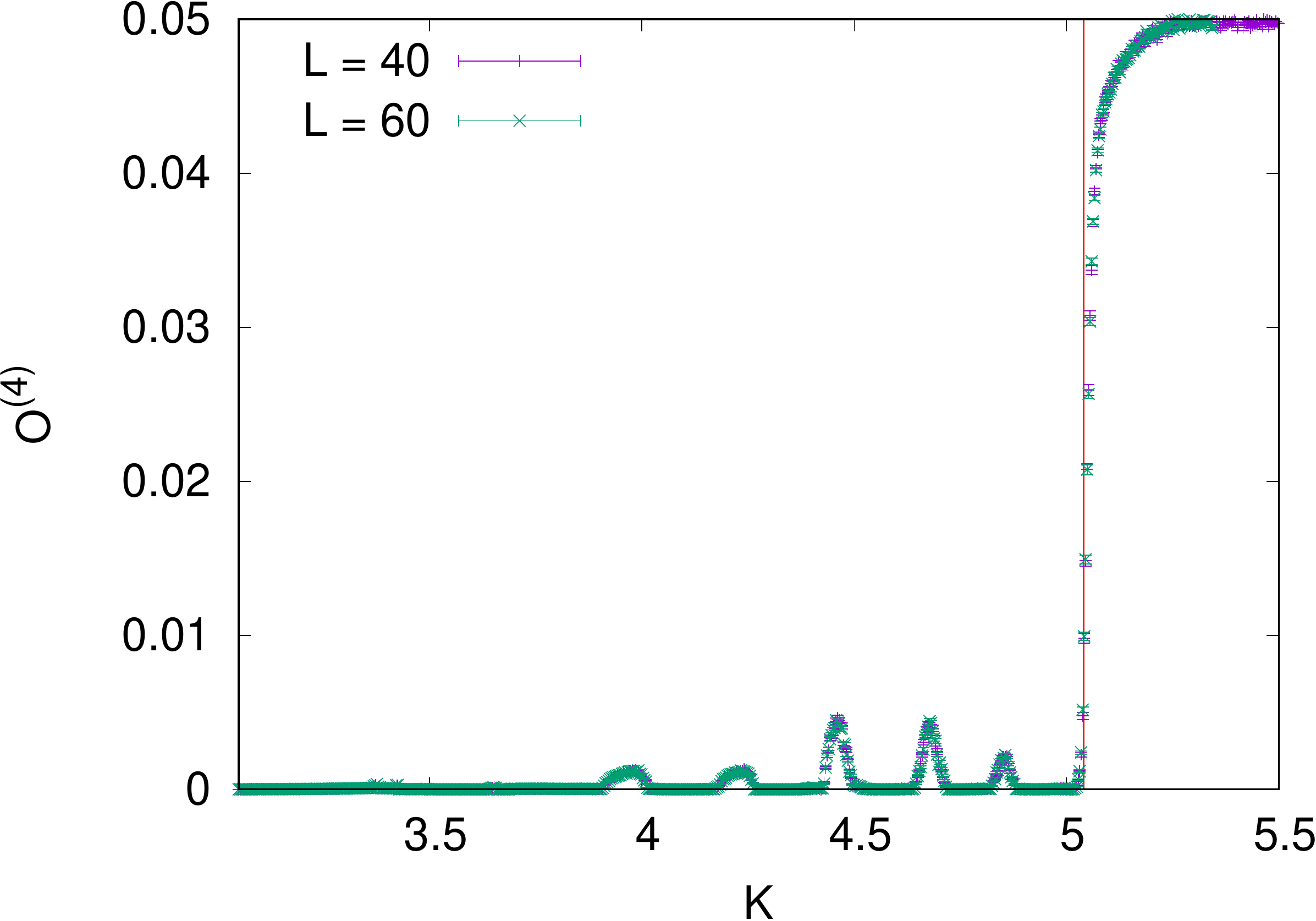}
    \end{tabular}
     \caption{Many rotors with $J=0.1$. (Panel a) Period-doubling order parameters versus $K$ for different values of $L$. (Panel b) Period $4$-tupling order parameter versus $K$ for different values of $L$. The vertical line marks the onset of chaos, as found with the LLE analysis. Numerical parameters $\mathcal{T}=4\cdot 10^4,\,n_0=10^3, N_{\rm r} = 10^3$.}\label{mop:fig}
\end{figure}

If we add the interactions we see that the period-doubling cascade is washed away, as we see in Fig.~\ref{mop:fig}. Focusing on the regular-dynamics regime, we see that the period-doubling response is dominant. There are still some small regions with period 4-tupling, but the response at $m=4$ is three orders of magnitude smaller that the one at $m=2$, making it quite negligible. Again in the chaotic regime there is a response at all frequencies due to chaos and we do not consider it. In the figure we choose a specific value of $J$, but we have checked that the thing is general. 

So, we essentially find confirmation of the results of~\cite{PhysRevA.41.1932}: whenever the dynamics is regular, and then makes sense speaking about period $m$-tupling, the dominant response is period doubling, with some negligible contribution at larger frequency. The fact that this contribution exists is probably due to the fact that our system has continuous local variables, in contrast with the discrete local variables of cellular automata~\cite{PhysRevA.41.1932}, and the known classical dissipative time crystals~\cite{zhuang2021absolutely,Pizzi_2021,PhysRevE.100.060105}.

We see in Fig.~\ref{mop:fig}(a) that $\mathcal{O}^{(2)}$ is vanishing for $K$ up to a threshold and then abruptly moves from 0. The threshold where this happens marks the line bounding from below the ``Spatiotemporal ordering'' regime in Fig.~\ref{phd:fig}. 
\section{Behavior of the kinetic energy}\label{nono:sec}
\begin{figure}
  \begin{tabular}{c}
    \includegraphics[width=80mm]{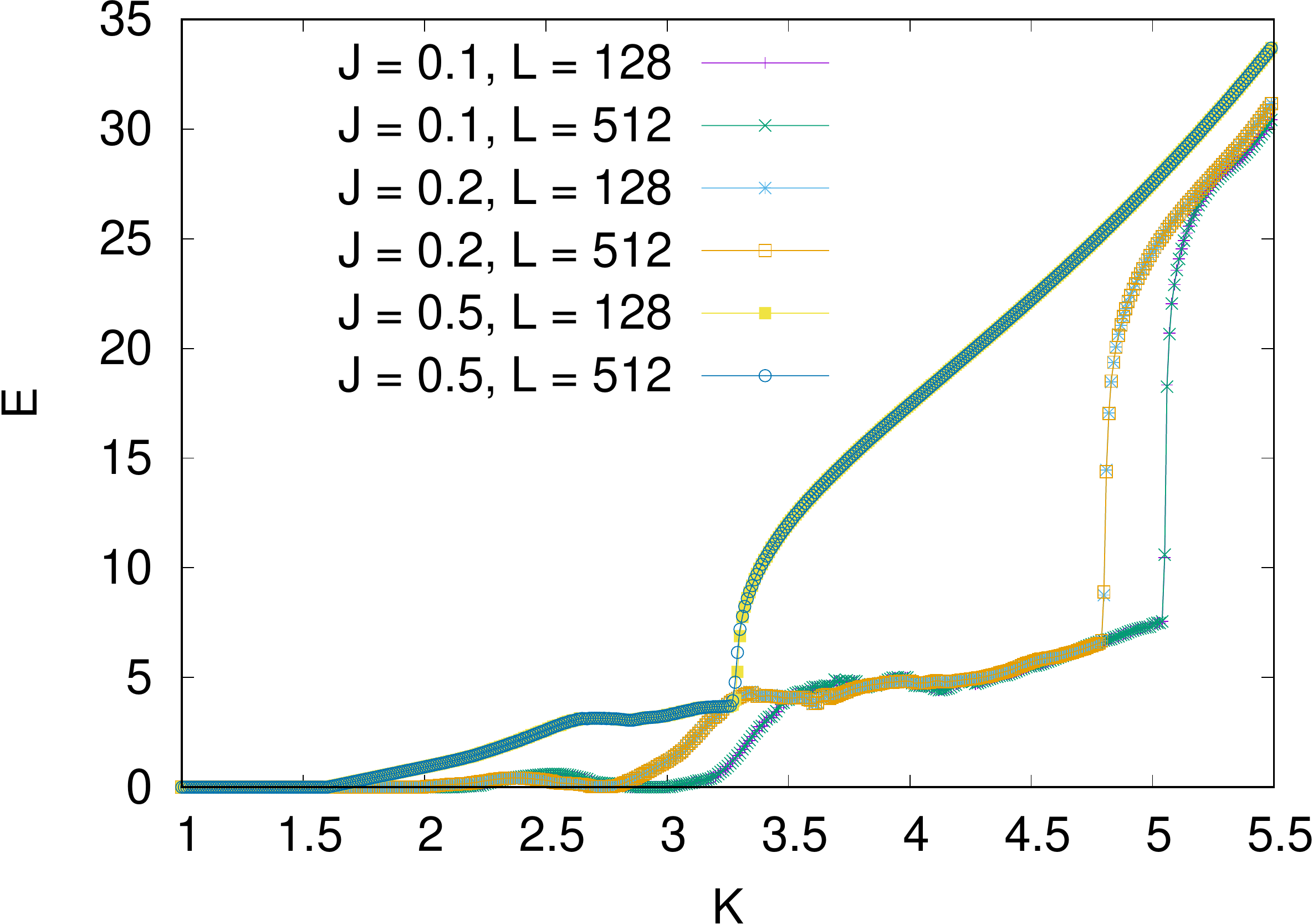}
  \end{tabular}
   \caption{Onsite kinetic energy averaged over time and initial-state realizations [Eq.~\eqref{E:eqn}] versus $K$, for different values of $J$ and $L$. Numerical parameters: $N_{\rm r}=10^3,\,n_0=10^3,\,\mathcal{T}=6\cdot 10^4$.}\label{epsq01:fig}
\end{figure}

Some properties of the patterns can be read in the kinetic energy per site defined as
\begin{equation}
  E(n) = \frac{1}{2L}\sum_j [p_j^{(n)}]^2\,.
\end{equation}
We consider its average over initial-state realizations and time
\begin{equation}\label{E:eqn}
  E = \frac{1}{n_0}\sum_{n=\mathcal{T}-n_0}^{\mathcal{T}}\overline{E(n)}\,,
\end{equation}
where $\mathcal{T}$ and $n_0$ are chosen so that convergence is reached. We see first of all that this quantity is finite, also in the chaotic regime (see Fig.~\ref{epsq01:fig}). This is an important difference compared with the Hamiltonian case, where chaos leads to an unbounded increase in time of energy. Moreover we see in this quantity the same features that we see in $\delta p$ (compare with Fig.~\ref{dpn:fig}). In particular it vanishes in the trivial regime, and shows a discontinuity in the derivative and a sudden increase at the onset of chaos. Therefore, also the kinetic energy can be used to find if there is pattern formation.
%

In Fig.~\ref{epsq01:fig} we see also that the energy per site is bounded and does not scale with the system size (it is actually size-independent for the values of $L$ we are considering). This is a key point to make possible an experimental realization with an array of Josephson junctions, as we are going to show in the next section.
\section{Proposal for experimental realization}\label{exp:sec}
\begin{figure}
  \begin{tabular}{c}
    \includegraphics[width=80mm]{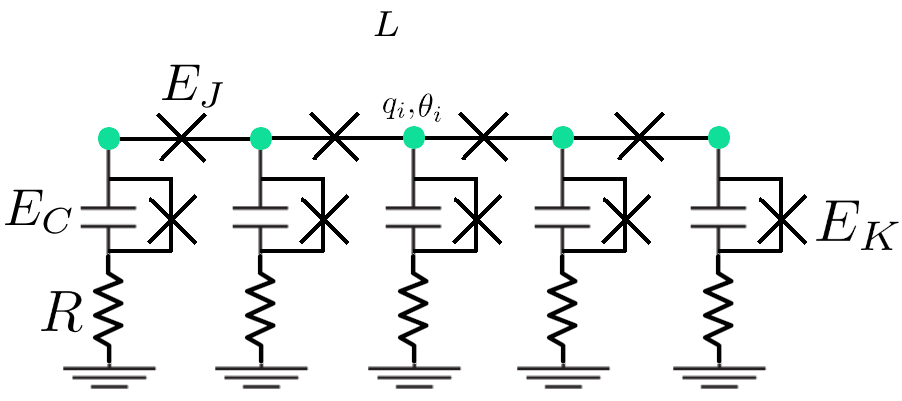}
  \end{tabular}
   \caption{Josephson junction array for the experimental realization of the model. Here we show open boundary conditions. In order to get periodic boundary conditions, one more junction $E_J$ is needed connecting the first and the last sites. }\label{array:fig}
\end{figure}

The Hamiltonian in Eq.~\eqref{hamiltone:eqn} can be realized by means of an array of Josephson junctions (see for instance~\cite{Leggett,tinkham:book,Devoret} for details). We show it in Fig.~\ref{array:fig}. With the crosses we mark SQUIDs, that's to say Josephson junctions whose energy can be tuned with the application of a magnetic flux. So, we have $E_J=E_J(\Phi)=E_J^{(0)}\cos(2\pi\Phi/\Phi_0)$ for the junctions in the upper row and $E_K=E_K(\Phi)=E_K^{(0)}\cos(2\pi\Phi/\Phi_0)$ for the junctions in the lower row, where $\Phi$ is the magnetic flux and $\Phi_0 = hc /(2e)$ the Cooper-pair flux quantum. For the junctions in the upper row we neglect the capacitance, for each of those in the lower row the capacitance is $C$, so that the corresponding charging energy is $E_C = e^2/C$. For each of the junction of the lower row the corresponding capacity is connected in parallel, and each of these parallel circuits is connected to the ground by means of a resistance $R$. {As in~\cite{Pino536,PhysRevLett.122.054102} we assume we can neglect the linear inductances that are always present in the wires.}

On each site (the green balls) we have two canonically conjugate dynamical variables, the gauge invariant superconducting phase $\theta_j$ and the charge $q_j$ (expressed in units of the electron charge $e$). If we assume that the superconducting pieces just above the resistances have all the same phase $\varphi$, up to a shift of the $\theta_j$, we can set the phases to 0 ($\varphi=0$). When there is no resistance, the circuit is described by the Hamiltonian
\begin{equation}\label{hammi:eqn}
  H = \sum_{j=1}^L\frac{1}{2}E_Cq_j^2 - E_J(\Phi)\cos(\theta_j-\theta_{j+1}) - E_K(\Phi)\cos(\theta_j)\,.
\end{equation}
{In Appendix~\ref{demo:sec} we show that the dynamics with no resistance is described by this Hamiltonian.} In the figure open boundary conditions are represented, but we can impose periodic boundary conditions by adding a junction $E_J$ connecting the first and the last sites. We apply the following time-periodic protocol:
\begin{enumerate}
 \item For a time $T_1$ the flux $\Phi$ is kept equal to $\Phi=\Phi_0/4$ (all the SQUIDs have vanishing Josephson energy and they are closed);
 \item For a time $T_2$  the flux $\Phi$ is kept equal to $\Phi=0$ (all the SQUIDs have maximum Josephson energy and they are open);
\end{enumerate}
%
Physically, this corresponds to a periodic protocol with period $T_1+T_2$ where a kick lasting $T_2$ is applied, because during the interval $T_1$ the Josephson energies of the SQUIDs vanish and there is no Josephson energy term acting. 

We consider the stroboscopic evolution, looking at what happens from just before one kick to just before the next, that's to say from time $t_n=n(T_1+T_2)+T_1$ to $t_{n+1} = (n+1)(T_1+T_2)+T_1$. The dynamics is provided by the canonical equations 
\begin{eqnarray}\label{cano:eqn}
  \frac{\hbar}{{2e}} \dot{q}_j&=-\partial_{\theta_j} H\,,\nonumber\\
  \frac{\hbar}{{2e}} \dot{\theta}_j&=\phantom{-}\partial_{q_j} H\,.
\end{eqnarray}
{(The coefficient in front of the derivative is immaterial and can be eliminated by appropriately rescaling the Hamiltonian or the canonical coordinates.)} We can solve these equations and, if the time of the kick is much smaller than $T_2\ll \min\left(\frac{\sqrt{E_J^{(0)}E_C}}{\hbar},\,\frac{\sqrt{E_J^{(0)}E_C}}{\hbar}\right)$, we can approximate the evolution from time $t_n=n(T_1+T_2)+T_1$ to $t_{n+1} = (n+1)(T_1+T_2)+T_1$ as a discrete map
\begin{align}\label{qt:eqn}
  \frac{\hbar}{2e} q_j(t_{n+1}) & = \frac{\hbar}{2e} q_j(t_{n}) -T_2 E_J^{(0)}(\sin(\theta_j-\theta_{j+1})+ \sin(\theta_j-\theta_{j-1})) \nonumber\\
    &- T_2 E_K^{(0)}\sin(\theta_j)\nonumber\\
  \frac{\hbar}{2e} \theta_j(t_{n+1}) & = \frac{\hbar}{2e} \theta_j(t_n) + T_2 E_C q_j(t_{n+1})\,.
\end{align}
Applying the change of variables $\theta_j^{(n)} = \theta_j(t_{n})$, $p_j^{(n)} = 2e\frac{T_2E_C}{\hbar}q_j(t_{n})$, and defining $J\equiv 4e^2T_2^2\frac{E_CE_J^{(0)}}{\hbar^2}$, $K\equiv 4e^2T_2^2\frac{E_CE_K^{(0)}}{\hbar^2}$ we get back the Hamiltonian mapping Eq.~\eqref{hami:eqn}.

In order to get the model with dissipation, we consider a nonvanishing resistance $R$.  About the presence of the resistances, we emphasize that they are quite realistic, because they are naturally present and the difficult thing is removing them, rather than adding. With the resistance, between one kick and the next, the junctions are switched off and the charges damp as in the $RC$ circuit, so that the map Eq.~\eqref{qt:eqn} becomes
\begin{align}
  \frac{\hbar}{2e} q_j(t_{n+1}) & = \nep^{-RCT_1}\big[\hbar q_j(t_{n}) -T_2 E_J^{(0)}(\sin(\theta_j-\theta_{j+1})\nonumber\\
    &+ \sin(\theta_j-\theta_{j-1})) - T_2 E_K^{(0)}\sin(\theta_j)\big]\nonumber\\
  \frac{\hbar}{2e} \theta_j(t_{n+1}) & = \hbar \theta_j(t_n) + E_C\left(\frac{1-\nep^{-RC T_1}}{RC}\right) q_j(t_{n+1})\,,
\end{align}
provided that $T_2\ll RC$. Applying the change of variables $\theta_j^{(n)} = \theta_j(t_{n})$, $p_j^{(n)} = 2e\frac{E_C}{\hbar}\left(\frac{1-\nep^{-RC T_1}}{RC}\right)q_j(t_{n})$, and defining 
\begin{align}
  \gamma&\equiv \nep^{-RC T_1}\nonumber\\
  J&\equiv 4e^2T_2\frac{E_CE_J^{(0)}}{\hbar^2}\nep^{-RCT_1}\left(\frac{1-\nep^{-RCT_1}}{RC}\right)\nonumber\\
  K&\equiv  4e^2T_2\frac{E_CE_K^{(0)}}{\hbar^2}\nep^{-RCT_1}\left(\frac{1-\nep^{-RCT_1}}{RC}\right)
\end{align}
we get back the dissipative mapping Eq.~\eqref{hami:eqn}.

There is an important consideration about the charging energy per site, given by $E_q(n) = \frac{E_C}{2L}\sum_{j=1}^L[q_j(t_{n+1})]^2$. This quantity is proportional to the kinetic energy per site we have discussed in Sec.~\ref{nono:sec}, and we have seen there that in the dissipative model it attains an asymptotic value order 1. By appropriately tuning the parameters, there is the possibility to keep this asymptotic value smaller than the superconducting gap of the system. In this way, the model obeys for long times the dynamics we have described here. 
In contrast with that, in case of Hamiltonian evolution, the energy at some point starts increasing in an unbounded way~\cite{PhysRevB.100.100302,PhysRevA.40.6130,Konishi_1990,PhysRevB.100.100302} and at some point it goes beyond the gap and superconductivity is lost.
\section{Conclusion}\label{conc:sec}
In conclusion we have studied a model of coupled kicked rotors with dissipation. In the case of a single rotor, the model reduces to the Zaslavsky map and shows a behavior strictly similar to the period-doubling route to chaos, that is so widespread in Nature. (Actually we see two parallel period-doubling cascades -- see Fig.~\ref{bd:fig}.

We focus on the momenta probed at discrete stroboscopic times (at each period of the driving) and consider the appropriate averages over many random initial conditions. Using some period $m$-tupling order parameters inspired to the time-crystal literature, we see that in the many-rotor case the period-doubling cascade disappears and one can essentially see only period doubling, with here and there much smaller contributions at period four times the driving. So, we find essentially confirmation of the findings of Ref.~\cite{PhysRevA.41.1932}, with the presence of the small period-doubling contributions probably related to the fact that this model has local continuous variables, in contrast with the discrete ones of cellular automata and classical time crystals.

The dynamics of this model is very rich. First of all we see that the period doubling always appears in association with the spontaneous formation of patterns, that are stable and persistent in time, and depend on the stroboscopic time with periodicity double than the one of the driving. Therefore the system breaks at the same time the discrete time translation symmetry and the space translation symmetry, therefore we talk about ``spatiotemporal ordering''. A similar phenomenon occurs in quantum Floquet time crystals, but the physics is different: Here an effect of the classical nonlinear dynamics, there a sort of quantum phase transition. Indeed, here we see a persistent spatiotemporal ordering already at finite sizes, while in time crystals it persists only in the thermodynamic limit.

{Most remarkably, the spatiotemporal order can be predicted analytically, by means of a linear stability analysis around the state with all vanishing momenta and coordinates. This analysis predicts an instability region, whose boundary coincides precisely with the lower boundary of the spatiotemporally ordered phase. Moreover  the eigenvalues of the stroboscopic dynamics are real and negative, so that the most unstable mode grows changing sign at any period, until this growth stops due to nonlinearities, and a period-doubled steady pattern develops. Another interesting finding is that the momentum of the most unstable mode is $k=\pi$, giving rise to a typical length scale double than the lattice spacing. We find that this prediction of the linear-stability analysis is valid for the fully-developed patterns in the spatiotemporally ordered regime, that have precisely this typical length scale. Therefore period doubling in space is associated to period doubling in time.}

There is also a phase where the system breaks only the space-translation symmetry and the patterns are stable and independent from the stroboscopic time. Inside this region of the phase space there is a smaller region that we call ``Weak patterning'' and where points with no patterns alternate with point with small-amplitude patterns, in a jagged and apparently fractal way.

Beyond that, there is the trivial regime, where the model relaxes to a uniform condition with vanishing momentum, and the chaotic regime, where the largest Lyapunov exponent is larger than 0, nearby trajectories in phase space diverge exponentially, and the dynamics is aperiodic and irregular in space and time.

The transition from different regimes can be seen in the properties of the patterns. For instance, the pattern typical length has a sudden drop moving from time-independent patterning to spatiotemporal ordering, and -- in a range of parameters -- shows a peak at the onset of chaos where it increases of one order of magnitude. { Inside the spatiotemporally ordered phase, near the lower phase boundary, we find also a typical length scale near the value $\lambda = 2$ predicted by linear stability analysis.}

Many transitions in the dynamics can be seen also in the behavior of the pattern amplitude, that anyway misses the threshold between time-independent patterning and spatiotemporal ordering, because changes in a continuous and regular way at this threshold. The same information given by the amplitude can be understood from the behavior of the asymptotic average kinetic energy per site. This quantity is finite, also in the chaotic regime, in contrast with the unbounded increase of energy in case of Hamiltonian chaos. Moreover, in our case, this quantity does not scale with the system size.

This is an important information for experimental realization. Indeed, we propose to realize this model in an array of SQUID Josephson junctions, and the fact that the system does not heat up above a threshold gives the opportunity to choose the parameters in such a way that the system stays superconducting, and our model is a good description thereof for long times.

About future developments, {it is first of all compelling to understand why linear stability analysis does not predict the time-independent patterns.} 
 Another possibility is studying the patterns in different geometries (for instance in a 2-dimensional lattice) and see if similar dynamical behaviors appear if one couples other systems showing the period doubling cascade (for instance the nonlinear electric circuits of~\cite{PhysRevLett.47.1349,PhysRevLett.48.714}). One might think also to use the methods of~\cite{dittrich} to quantize the model of coupled dissipative rotors and, of course, realize the experimental proposal of Sec.~\ref{exp:sec}.
\acknowledgements{I thank A.~Delmonte, R.~Fazio, D.~Mukamel, G.~Passarelli, S.~Ruffo, and G.~E. Santoro for interesting discussions, and V.~Russomanno for having drawn my attention on Ref.~\cite{baba:book} and the problems of chaos and period-doubling cascade. I acknowledge P.~Lucignano for the access to the qmat machine where most of the numerics for this project was performed, and thank the ICTP for the warm hospitality received (under ERC Project 101053159 -- RAVE) during the completion of this work. I acknowledge financial support from PNRR MUR Project PE0000023-NQSTI.
\appendix
\section{Validiy of the Hamiltonian Eq.~\eqref{hammi:eqn} in the case without resistance}\label{demo:sec}
In order to show the validity of the Hamiltonian Eq.~\eqref{hammi:eqn}, let us write the Kirchhoff's equations for the circuit in Fig.~\ref{array:fig} without resistance. In order to do that, let us call $I_{j}$ the current moving from the island $j$ to the island $j+1$, $I_{j-1}$ the current from the island $j-1$ to the island $j$, $I_{S\,j}$ the current from the superconducting island $j$ to the ground (see Fig.~\ref{arrayolo:fig}).
\begin{figure}
  \begin{tabular}{c}
    \includegraphics[width=40mm]{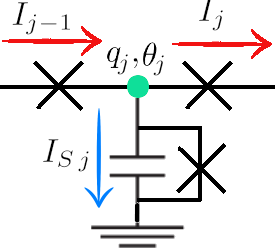}
  \end{tabular}
   \caption{Currents at the node where the island $j$ is. }\label{arrayolo:fig}
\end{figure}
Using the first Kirchhoff law at the node of the island we get 
$$
  I_{j-1} = I_{j} + I_{S\,j}\,.
$$
Using the definition of Josephson current we get $I_{j-1} = \frac{2e}{\hbar}E_J(\Phi)\sin(\theta_{j}-\theta_{j-1})$,  $I_{j} = \frac{2e}{\hbar}E_J(\Phi)\sin(\theta_{j+1}-\theta_{j})$. Using the Kirchhoff's law at the bifurcation to the parallel of the Josephson junction below and the capacitor, we get
$$
  I_{S\,j} = - \dot{q}_j -  \frac{2e}{\hbar}E_K(\Phi)\sin(\theta_{j})\,.
$$
Putting all these formulae together we get
\begin{equation}
  \dot{q}_j = \frac{2e}{\hbar}\left[E_J\sin(\theta_{j+1}-\theta_j)-E_J\sin(\theta_j-\theta_{j+1}) - E_K\sin\theta_j\right]\,.
\end{equation}
It is easy to check that this formula is provided by the canonical equation $\dot{q}_j = -\frac{2e}{\hbar}\partial_{\theta_j}H$ [see Eq.~\eqref{cano:eqn}], with $H$ given by Eq.~\eqref{hammi:eqn}.

We can also easily write the equations for the derivative of the phase. Applying the known formula~\cite{tinkham:book} at the junction below, we get 
$$
  \hbar \dot{\theta}_j = 2e\Delta V_j = 2e\frac{q_j}{C}\,,
$$
 where $\Delta V_j$ is the voltage difference between the island $j$ and the ground. The point is that $\theta_j$ is the phase difference across the Josephson junction below, and the voltage difference across this junction is provided by the voltage difference across the capacitor, $\Delta V_j = \frac{q_j}{C}$. It is easy to show that this equation for the derivative of $\theta_j$ is provided by the canonical equation $\dot{\theta}_j = \frac{2e}{\hbar}\partial_{q_j}H$ [see Eqs.~\eqref{cano:eqn}], with $H$ provided by Eq.~\eqref{hammi:eqn}.
%
%
%
%
\end{document}